\input amstex
\documentstyle{amsppt}
\magnification=1200

\loadbold

\TagsOnRight
\NoRunningHeads

\define\al{\alpha}

\define\ga{\gamma}
\define\Ga{\Gamma}
\define\de{\delta}
\define\ka{\kappa}
\define\la{\lambda}
\define\Om{\Omega}
\define\om{\omega}
\define\th{\theta}
\define\ze{\zeta}
\define\ep{\varepsilon}

\define\C{\Bbb C}
\define\R{\Bbb R}
\define\E{\Bbb E}
\define\SS{\Bbb S}

\define\T{\Bbb T}

\define\diag{\operatorname{diag}}
\define\tr{\operatorname{tr}}
\define\Hreg{H_{\operatorname{reg}}}
\define\reg{{\operatorname{reg}}}
\define\Orbb{\operatorname{Orb}}
\define\Conf{\operatorname{Conf}}
\define\const{\operatorname{const}}
\define\Arg{\operatorname{Arg}}
\define\sgn{\operatorname{sgn}}
\define\ex{\operatorname{ex}}

\define\tht{\thetag}
\define\wt{\widetilde}

\define\ms{m^{(s)}}
\define\msN{m^{(s,N)}}

\define\M{\operatorname{Mat}(N,\C)}
\define\Mplus{\M_+}
\define\MM{\frak M}

\define\p{\boldsymbol p}
\define\K{\boldsymbol K}
\define\r{\boldsymbol \rho}

\topmatter

\author Alexei Borodin and Grigori Olshanski
\endauthor
\title Infinite random matrices and ergodic measures
\endtitle

\abstract
We introduce and study a 2--parameter family of unitarily invariant
probability measures on the space of infinite Hermitian matrices. 
We show that the decomposition of a measure from this family on ergodic
components is described by a determinantal point process on the real line. The
correlation kernel for this process is explicitly computed. 

At certain values of parameters the kernel turns into the well--known sine
kernel which describes the local correlation in Circular and Gaussian Unitary
Ensembles. Thus, the random point configuration of the sine process is
interpreted as the random set of ``eigenvalues'' of infinite
Hermitian matrices distributed according to the corresponding measure.   

\endabstract

\date Preliminary version. October 11, 2000 \enddate

\toc
\widestnumber\head{???}
\head {} Introduction \endhead
\head 1. The pseudo--Jacobi ensemble \endhead
\head 2. The scaling limit of the correlation functions \endhead
\head 3. The Hua--Pickrell measures \endhead
\head 4. Ergodic measures \endhead
\head 5. Approximation of spectral measures \endhead
\head 6. The main result \endhead
\head 7. Vanishing of the parameter $\ga_2$ \endhead
\head 8. Remarks and problems \endhead
\head 9. Appendix: Existence and uniqueness of decomposition on
ergodic components \endhead 
\head {} References \endhead
\endtoc

\endtopmatter

\document

\head Introduction \endhead

We first introduce some basic notions, and then describe the main 
results of the paper. 

\subhead Random point configurations and correlation functions
\endsubhead
Let $\frak X$ be a locally compact space. A {\it locally finite point
configuration\/} in 
$\frak X$ is a finite or countably infinite collection of points in
$\frak X$, also called {\it particles,\/} such that any compact set contains
finitely many particles. The ordering of the particles is
unessential. For the sake of brevity, we will omit the adjective
`locally finite'. A {\it point process\/} on $\frak X$ is a
probability measure on the space $\Conf(\frak X)$ of point
configurations. Given a point process, we can speak
about the {\it random\/} point configuration. The $n$th {\it
correlation measure\/} of a 
point process ($n=1,2,\dots$) is a symmetric measure $\rho_n$ on
$\frak X^n$, which is determined by the relation
$$
\langle \rho_n, F\rangle=
\Bbb E\left(\sum F(x_1,\dots,x_n)\right), \tag0.1
$$
where $F$ is a compactly supported test function on $\frak X^n$,
$\Bbb E$ is the symbol of expectation, and the summation is taken
over all ordered $n$-tuples of particles chosen from the random
point configuration. The $n$th {\it correlation function\/} is the density of
$\rho_n$ with respect to the $n$th power of a certain reference
measure on $\frak X$. Usually, the reference measure is the Lebesgue
measure. The first correlation function is also called
the {\it density function}. See \cite{Len}, \cite{DVJ, Ch. 5}\footnote{In the book \cite{DVJ} the correlation measures are called the ``factorial moment measures''.},
\cite{So}. 

\subhead The Dyson circular unitary ensemble \endsubhead
Let $\Bbb T\subset\C$ be the unit circle and $\Bbb T^N/S(N)$ be the set
of orbits of the symmetric group $S(N)$ of degree $N$ acting on the
torus $\Bbb T^N$, where $N=1,2,\dots$. Consider the following
probability measure on $\Bbb T^N/S(N)$:
$$
\const\cdot\prod_{1\le j<k\le N}|u_j-u_k|^2\,\prod_{j=1}^N
d\varphi_j, \qquad u_j=e^{2\pi i\varphi_j}\in\Bbb T\,,
\quad \varphi_j\in[-\tfrac12,\tfrac12], \tag0.2
$$
where $\const$ is the normalizing factor, $i=\sqrt{-1}$. 
This measure defines a point process on $\frak X=\Bbb T$ living on
the $N$-point configurations, which is called the $N$th {\it Dyson
circular unitary ensemble\/} or simply the Dyson ensemble for short. Note that the Dyson ensemble is invariant under rotations of $\Bbb T$.

Let $U(N)$ be the group of $N\times N$ unitary matrices. Consider
the natural projection $U(N)\to\Bbb T^N/S(N)$ assigning to a matrix
$U\in U(N)$ the collection of its eigenvalues. Note that the fibers
of this projection are exactly the conjugacy classes of the group
$U(N)$. The measure \tht{0.2} coincides with the pushforward of the
normalized Haar measure on $U(N)$ under this projection. In other
terms, \tht{0.2} is the {\it radial part\/} of the Haar measure. It follows
that the Dyson ensemble is formed by spectra of random unitary
matrices $U\in U(N)$ distributed according to the Haar measure. See
\cite{Dys}, \cite{Me}. 

\subhead The sine process \endsubhead
This is a translationally invariant point process on $\frak X=\R$.
Its correlation functions (with respect to the Lebesgue measure on
$\R$) are given by 
$$
\rho_n(y_1,\dots,y_n)=
\det\left[\frac{\sin(\pi(y_j-y_k))}{\pi(y_j-y_k)}\right]_{j,k=1}^n\,,
\qquad n=1,2,\dots, \quad y_1,\dots,y_n\in\R. \tag0.3
$$
The function $\frac{\sin(\pi(y-y'))}{\pi(y-y')}$ on $\R\times\R$ is
called the {\it sine kernel.} 

The correlation functions of the sine process can be obtained from
the correlation functions of the $N$th Dyson ensemble by the following
scaling limit as $N\to\infty$. Fix an arbitrary point $u_0\in\Bbb T$
and rescale the angular coordinate $\varphi$ about the point $u_0$ by
writing $u=u_0e^{2\pi i y/N}$. Then, for any fixed $n$, the $n$th
correlation function of the $N$th Dyson ensemble, expressed in terms
of the $y$--variables, converges, as $N\to\infty$, to the function
\tht{0.3}. See \cite{Dys}, \cite{Me}.  

\subhead A substitute of the Haar measure \endsubhead
A natural question is whether the sine process can be interpreted as
a radial part of an infinite--dimensional analog of the Haar measure.
In this paper we suggest such an interpretation. 

It is convenient to pass from unitary matrices to Hermitian matrices.
Let $H(N)$ be the linear space of $N\times N$ complex Hermitian
matrices. Consider the Cayley transform
$$
H(N)\ni X\mapsto U=\frac{i-X}{i+X}\in U(N), \qquad N=1,2,\dots \tag0.4
$$
The map \tht{0.4} is one--to--one, and the complement of its image in
$U(N)$ is a negligible set. Thus, we can transfer the normalized
Haar measure from $U(N)$ to $H(N)$. The result has the following form
$$
\const\cdot\det(1+X^2)^{-N}\times(\text{the Lebesgue measure}).
\tag0.5 
$$

Let $H$ be the space of all infinite Hermitian matrices
$X=[X_{jk}]_{j,k=1}^\infty$. A remarkable fact is that the measures
\tht{0.5} with different values of $N$ are consistent with natural
projections $H(N)\to H(N-1)$ and, therefore, determine a probability
measure $m$ on $H$. We view $m$ as a substitute of the Haar measure on
$U(N)$ for $N=\infty$. 

\subhead Ergodic measures \endsubhead
Assume that we have a group acting on a Borel space. An invariant
probability Borel measure is called {\it ergodic\/} if any invariant
mod 0 set has measure 0 or 1. Ergodic measures coincide with extreme
points of the convex set of all invariant probability measures, see
\cite{Ph}. For continuous actions of compact groups ergodic
measures are exactly orbital measures, i.e., invariant probability
measures supported by orbits. According to the general philosophy of
the ergodic theory, the concept of ergodic measure is a right
generalization of that of orbital measure. 

We are interested in a special situation when the space is $H$ and the
group is an infinite--dimensional version $U(\infty)$ of the groups
$U(N)$. By definition, $U(\infty)$ is the union of the groups $U(N)$.
Its elements are infinite unitary matrices $[U_{jk}]_{j,k=1}^\infty$
with finitely many entries $U_{jk}$ not equal to $\de_{jk}$. The
group $U(\infty)$ acts on the space $H$ by conjugations. 

Consider the space $\Om$ whose elements $\om$ are given by 2 infinite
sequences 
$$
\al^+_1\ge\al^+_2\ge\dots\ge0, \quad 
\al^-_1\ge\al^-_2\ge\dots\ge0, \qquad
\text{where}\quad \sum_{j=1}^\infty(\al^+_j)^2+
\sum_{j=1}^\infty(\al^-_j)^2<\infty, \tag0.6
$$
and 2 extra real parameters $\ga_1$, $\ga_2$, where $\ga_2\ge0$. 

It is known that the ergodic measures on $H$ can be parametrized by the
points $\om\in\Om$. We consider $\Om$ as a substitute of the space 
$\Bbb T^N/S(N)$ for $N=\infty$.  

Let us explain the asymptotic meaning of the parameters $\al^\pm_j,
\ga_1,\ga_2$. According to a general result, each ergodic measure $M$
on $H$ can be approximated by a sequence 
$\{M^{(N)}\mid N=1,2,\dots\}$, where $M^{(N)}$ is an orbital measure
on $H(N)$ with respect to the action of $U(N)$ by conjugations. Any
such measure $M^{(N)}$ is specified by a collection $\la^{(N)}$ of
eigenvalues. Then the parameters of $\om$ describe the asymptotic
behavior of $\la^{(N)}$ as $N\to\infty$:
$$
\gathered
\la^{(N)}=(\la^{(N)}_1\ge\dots\ge\la^{(N)}_N)
\sim(N\al^+_1,N\al^+_2,\dots,-N\al^-_2,-N\al^-_1), \\
\frac{\la^{(N)}_1+\dots+\la^{(N)}_N}N\to\ga_1\,,\\
\frac{(\la^{(N)}_1)^2+\dots+(\la^{(N)}_N)^2}{N^2}\to
\ga_2+(\al^+_1)^2+(\al^+_2)^2+\dots+(\al^-_1)^2+(\al^-_2)^2+\dots.
\endgathered \tag0.7
$$

For more detail, see \cite{Pi2}, \cite{OV}, and references therein.

\subhead From spectral measures to point processes \endsubhead 
It can be proved that any $U(\infty)$-invariant probability measure
on $H$ can be decomposed on ergodic components. I.e., it can be written
as a continual convex combination of ergodic measures. This
decomposition is unique, we call it the {\it spectral 
decomposition.\/} It is determined by a probability measure on $\Om$,
which we call the {\it spectral measure\/} of the initial invariant measure. 

We map the space $\Om$ to the space $\Conf(\R^*)$ of point
configurations on the punctured real line $\R^*=\R\setminus\{0\}$ as
follows: 
$$
\Om\ni\om=(\{\al^+_j\},\{\al^-_j\},\ga_1,\ga_2)\mapsto
C=(-\al^-_1,-\al^-_2,\dots,\al^+_2,\al^+_1)\in\Conf(\R^*), \tag0.8
$$
where we omit possible zeros among the numbers $\al^\pm_j$. The map
\tht{0.8} transforms any spectral measure (which is a probability
measure on $\Om$) to a point process on $\R^*$. This makes it
possible to describe spectral measures in terms of the correlation
functions. However, the map \tht{0.8} ignores the parameters $\ga_1,\ga_2$.

Note that each configuration $C\in\Conf(\R^*)$ of the form \tht{0.8}
is contained in a sufficiently large interval $|x|\le\const$. It
follows that $C^{-1}$ (the image of $C$ under the inversion map
$x\mapsto1/x$) is a well--defined configuration on the whole line
$\R$. 

\subhead An interpretation of the sine process \endsubhead
Applying the procedure described above to the measure $m$ on $H$ we
prove the following result.

\proclaim{Theorem I} Let $P$ be the spectral measure of the
$U(\infty)$-invariant measure $m$ and let $\Cal P$ be the
corresponding point process on $\R^*$. Then the point process on $\R$
obtained from $\Cal P$ under the transform $x\mapsto y=-\frac1{\pi x}$
coincides with the sine process.
\endproclaim

A lucid explanation of this result follows from the comparison of two
approximation procedures: that for the correlation functions of the
sine process and that for the ergodic measures. Indeed, the
eigenvalues in \tht{0.7} grow linearly in $N$, so that we rescale them
according to the rule $\la=Nx$. Under the Cayley transform
$u=\frac{i-\la}{i+\la}$ the scaling takes the form
$$
u=\frac{i-Nx}{i+Nx}=
-1+\frac{2i}{Nx}+O\left(\frac1{N^2}\right)=
(-1)e^{2\pi iy/N}+O\left(\frac1{N^2}\right), \qquad
y=-\frac1{\pi x}, \tag0.9
$$
which means that the variable $y$ is consistent with the scaling of the 
Dyson ensemble near the point $u_0=-1$. 

Thus, the statement of Theorem I is not surprising. However, the
justification of the formal limit transition made on the level of
correlation functions requires certain efforts.

Note also that dividing the eigenvalues $\la\in\R$ by $N$ corresponds
in terms of $u=\frac{i-\la}{i+\la}$ to the fractional--linear
transformation of $\Bbb T$ of the form
$$
u\mapsto\frac{(N+1)u+(N-1)}{(N-1)u+(N+1)}\,. \tag0.10
$$
This transformation has two fixed points, $+1$ and $-1$. Near the point
$-1$ it looks as the expansion by the factor of $N$ while near
the point $+1$ it looks as the contraction by the factor of $N$. Using
\tht{0.10} as a scaling transformation one can define a scaling limit
for the correlation functions of the Dyson ensembles staying on the
circle $\Bbb T$.

Theorem I is complemented by 

\proclaim{Theorem II} The spectral measure $P$ of the measure $m$ is
concentrated on the subset $\{\om\in\Om\mid \ga_2=0\}$.
\endproclaim

Thus, the parameter $\ga_2$ (which is ignored by the map \tht{0.8})
is actually irrelevant for the measure $m$. In a certain sense, this
means that the measure $m$ does not involve Gaussian
components (see \S4 about the connection of the parameter $\ga_2$
with Gaussian measures). 

\subhead A generalization: the main result \endsubhead
Let $s\in\C$, $\Re s>-\frac12$ be a parameter. Consider the following
probability measure on $\Bbb T^N/S(N)$:
$$
\gathered
\const\cdot\prod_{1\le j<k\le N}|u_j-u_k|^2\,\prod_{j=1}^N
(1+u_j)^{\bar s}(1+\bar u_j)^s d\varphi_j,\\
u_j=e^{2\pi i\varphi_j}\in\Bbb T\,,\quad 
\varphi_j\in[-\tfrac12,\tfrac12]. 
\endgathered
\tag0.11
$$ 
When $s=0$, we get \tht{0.2}. Thus, this is a deformation of the
measure \tht{0.2} depending on two real parameters, $\Re s$ and 
$\Im s$. The measure \tht{0.11} is the radial part of the 
probability measure on $U(N)$ of the form 
$$
\const\cdot\det((1+U)^{\bar s}) \det((1+U^{-1})^s)\times
\text{(the Haar measure on $U(N)$)}. \tag0.12
$$
Transferring the measure \tht{0.12} from the group $U(N)$ to the
space $H(N)$ by means of the Cayley transform \tht{0.4} we get the
following measure on $H(N)$, which is a deformation of the measure
\tht{0.5}: 
$$
\const\cdot\det((1+iX)^{-s-N})\det((1-iX)^{-\bar s-N})
\times(\text{the Lebesgue measure on $H(N)$}).
\tag0.13 
$$
When $s$ is real, the expression \tht{0.13} takes a simpler form:
$$
\gathered
\const\cdot\det((1+X^2)^{-s-N})
\times(\text{the Lebesgue measure on $H(N)$}),\\ s\in\R, \quad
s>-\tfrac12\,.
\endgathered 
\tag0.14 
$$

Again, it turns out that the measures \tht{0.13} are consistent with
the projections $H(N)\to H(N-1)$ and determine a
$U(\infty)$-invariant probability measure 
on the space $H$. We denote it by $m^{(s)}$. Note that $m^{(0)}=m$.

To our knowledge, the finite--dimensional measures \tht{0.14} were
first studied by Hua. He calculated the normalizing constant
factor in \tht{0.14} by a recurrence in $N$, and his argument proves
the consistence property (although he did not state it explicitly),
see \cite{Hua, Theorem 2.1.5}.
Much later Pickrell \cite{Pi1} considered analogs of the measures
\tht{0.12} and \tht{0.13} (with real $s$), which live on complex
Grassmannians and on the spaces of all complex matrices,
respectively. He proved the consistence property and considered
the analogs of the measures $\ms$ on the space of all complex
matrices of infinite order. His paper also contains a few other important
ideas and results. Apparently, Pickrell was unaware about Hua's
work. Note also Shimomura's paper \cite{Shim}, where an analog of the
measure $m^{(0)}$ for the infinite orthogonal group was constructed
(more general measures depending on a parameter are not discussed in
\cite{Shim}). The possibility of introducing a complex parameter (in
the case of Hermitian matrices) was discovered by Neretin
\cite{Ner2}. He also examined further generalizations of the measures
$\ms$.   

We propose to call the measures $\ms$ the {\it Hua--Pickrell measures.}

\proclaim{Theorem III} The Hua--Pickrell measures $m^{(s)}$ on $H$
are pairwise disjoint. I.e., for any two different values $s'$, $s''$ of the
parameter there exist two disjoint Borel subsets in $H$ supporting
$m^{(s')}$ and $m^{(s'')}$, respectively.
\endproclaim

The next claim is the main result of the paper.

\proclaim{Theorem IV} Let $P^{(s)}$ be the spectral measure of
a Hua--Pickrell measure $m^{(s)}$. The corresponding point process
$\Cal P^{(s)}$ on $\R^*$ can be described in terms of its correlation
functions. They have the determinantal form
$$
\rho^{(s)}_n(x_1,\dots,x_n)=\det[K^{(s)}(x_j,x_k)]_{j,k=1}^n\,, \tag0.15
$$
where $K^{(s)}(x,x')$ is a certain kernel on $\R^*\times\R^*$ which
can be expressed through the confluent hypergeometric
function or, for real values of $s$, through the Bessel function.
\endproclaim

See Theorem 2.1 below for explicit expressions for the kernel. As in Theorem I, one
can use the transformation $C\mapsto C^{-1}$ to pass from $\R^*$
to $\R$. 

\subhead Pseudo--Jacobi polynomials \endsubhead
The proof of Theorem IV, as that of Theorem I, goes in three steps:
the calculation of the correlation functions for the
finite--dimensional measures \tht{0.13}, the scaling limit transition
as $N\to\infty$, and a justification.  However, the first step, as
compared with the case of the Dyson ensemble, is more involved. We
show that the correlation functions are expressed through the
Christoffel--Darboux kernel for the so--called pseudo--Jacobi
polynomials. This family of orthogonal polynomials, which is not
widely known, has interesting features. It is defined by a weight
function on $\R$ possessing only finitely many moments, so that the
system of orthogonal polynomials is finite.

\subhead Organization of the paper \endsubhead
In \S1 we introduce the pseudo--Jacobi ensemble and obtain its correlation
functions.  In \S2 we compute the scaling limit of these correlation functions
as the number of particles goes to infinity. The limit correlation functions
are given by a determinantal formula and we  write down the correlation kernel
explicitly. In \S3 we define the Hua--Pickrell measures $m^{(s)}$ and show
that they are pairwise disjoint. \S4 provides a brief summary of known results
about the ergodic $U(\infty)$-invariant probability measures on $H$. In \S5 we
show that the spectral measure for any $U(\infty)$-invariant probability
measure $M$ on $H$ can be approximated by finite-dimensional projections of
$M$. \S6 contains the proof of our main result (Theorem IV above). In \S7 we
prove that the sine process has no Gaussian component (Theorem II above). \S8
contains remarks concerning the connections of our work with other
subjects as well as several open problems. \S9 is an appendix where we prove
the existence and uniqueness of the decomposition of $U(\infty)$-invariant
probability measures on $H$ on ergodic measures. 

\subhead Acknowledgment \endsubhead
At various stages of the work we discussed the subject with Sergei Kerov,
Yuri Neretin, and Anatoly Vershik. We are grateful to them for
valuable remarks. The second author (G.~O.) was supported by the
Russian Foundation for Basic Research, grant 98--01--00303.

\head 1. The pseudo--Jacobi ensemble \endhead

In this section we define the pseudo--Jacobi ensemble and compute
its correlation functions. 

Consider the radial part of the Haar measure on $U(N)$ which
determines the Dyson ensemble, see \tht{0.2}. Under the inverse Cayley
transform $\T\to\R$ which takes $u\in\T$ to $x=i\,\frac{1-u}{1+u}\in\R$,
the measure \tht{0.2} turns into the following 
measure on $\R^N/S(N)=\Conf_N(\R)$, the set of $N$-point configurations on
$\R$: 
$$
\const \,\prod_{1\le j<k\le N}(x_j-x_k)^2\cdot
\prod_{j=1}^N(1+x_j^2)^{-N}dx_j\,. \tag1.1
$$

More generally, let $s$ be a complex parameter. We introduce the
following deformation of the measure \tht{1.1} depending on $s$:
$$
\gathered
\const \,\prod_{1\le j<k\le N}(x_j-x_k)^2\cdot
\prod_{j=1}^N(1+ix_j)^{-s-N}(1-ix_j)^{-\bar s-N}dx_j\\
=\const\, \prod_{j=1}^N(1+x_j^2)^{-\Re s-N} 
e^{2\Im s\Arg(1+ix_j)}dx_j\,. 
\endgathered
\tag1.2
$$
Here we assume that the function $\Arg(\dots)$ takes values in
$(-\pi,\pi)$ (actually,
$\Arg(1+ix_j)\in(-\tfrac{\pi}2,\tfrac{\pi}2)$). 

\proclaim{Proposition 1.1} The measure \tht{1.2} is finite provided
that $\Re s>-\tfrac12$.  
\endproclaim

\demo{Proof} This follows from the estimate
$$
(1+x^2)^{-\Re s-N} e^{2\Im s\Arg(1+ix)}\asymp|x|^{-2\Re s-2N}, \qquad
x\in\R, \quad |x|\gg0, \tag1.3
$$
and the fact that the expansion of $\prod_{1\le j<k\le N}(x_j-x_k)^2$
involves only monomials of degree less or equal to $2N-2$ in each
variable. \qed
\enddemo

Henceforth we assume the condition $\Re s>-\frac12$ to be satisfied,
and we choose the normalizing constant in \tht{1.2} in such a way that
\tht{1.2} defines a probability measure. About the case $\Re s\le-\frac12$
see \S8 below. 

Note that \tht{1.2} corresponds, via the Cayley transform, to
the measure \tht{0.11}. 

For real values of the parameter $s$ the expression \tht{1.2} takes a
simpler form
$$
\const \,\prod_{1\le j<k\le N}(x_j-x_k)^2\cdot
\prod_{j=1}^N(1+x^2_j)^{-s-N}, \qquad s\in \R.
$$

Our aim is to compute the correlation functions of the measure
\tht{1.2}. We remark that \tht{1.2} is an orthogonal polynomial
ensemble (see \cite{Me}, \cite{NW}) corresponding to the weight function
$$
\phi(x)=(1+ix)^{-s-N}(1-ix)^{-\bar s-N}
=(1+x^2)^{-\Re s-N} e^{2\Im s\Arg(1+ix)}, \qquad x\in\R. \tag1.4
$$
We call it the $N$th {\it pseudo--Jacobi ensemble.\/} The reason for
this term is explained below. For generalities about orthogonal
polynomial ensembles, see, e.g., \cite{Me}, \cite{NW}.
 
Let $\p_0\equiv1,\p_1,\p_2,\dots$ denote the monic orthogonal
polynomials on $\R$ associated with the weight function \tht{1.4}.
Since for any $s$, $\phi(x)$ has only finitely many moments, this
system of orthogonal polynomials is finite. Specifically, it
follows from \tht{1.3} that the polynomial $\p_m(x)$ exists if 
$m<\Re s+N-\frac12$.  

According to a well--known general principle (see, e.g., \cite{Me}),
the correlation functions in question
are given by determinantal formulas involving the
Christoffel--Darboux kernel 
$$
\sum_{m=0}^{N-1}\frac{\p_m(x')\p_m(x'')}{\Vert \p_m\Vert^2}.
\tag1.5
$$
By the assumption $\Re s>-\frac12$, the polynomials up to the order
$m=N-1$ exist, so that this kernel makes sense.

The orthogonal polynomials $\p_m$ are known. They were introduced by
V.~Romanovski in 1929, see \cite{Ro}, and studied recently by
P.~A.~Lesky \cite{Les1, \S5}, \cite{Les2, \S1.4}.  They are called the
{\it pseudo-Jacobi polynomials,\/} which explains the choice of the
name for the ensemble \tht{1.2}.  

Let
$$
{}_2F_1\left[ \matrix a,\, b\\ c\endmatrix\Biggr|z\right]=
\sum_{n=0}^\infty \frac{a(a+a)\dots(a+n-1)\cdot b(b+1)\dots(b+n-1)}
{c(c+1)\dots(c+n-1)\cdot n!}\,z^n
$$
denote the Gauss hypergeometric function. 

\proclaim{Proposition 1.2} Let $m<\Re s+n-\frac12$, so that the $m$th
monic orthogonal polynomial $\p_m$ with the weight function \tht{1.4}
exists. Then it is given by the explicit formula 
$$
\p_m(x)=(x-i)^m{}_2F_1\left[ \matrix -m,\, s+N-m\\ 
2\Re s+2N-2m\endmatrix\Biggr|\frac 2{1+ix}\right]. \tag1.6
$$
and its norm is given by
$$
\multline
\Vert \p_m(x)\Vert^2=\int_{-\infty}^\infty \p_m^2(x)\phi(x)dx\\ =
\frac{\pi\,2^{-2\Re s}}{2^{2(N-m-1)}}\,\Gamma\bmatrix
 2\Re s+2(N-m)-1,\, 2\Re s+2(N-m),\,m+1\\s+N-m,\,\bar{s}+N-m,\,2\Re s+2N-m
\endbmatrix\,,
\endmultline \tag1.7
$$
where we use the notation
$$
\Gamma \bmatrix a,\, b,\,\dots\\ c, \, d, \dots\endbmatrix=
\frac{\Gamma(a)\Gamma(b)\dots}{\Gamma(c)\Gamma(d)\dots}\,.
$$
\endproclaim

\demo{Proof} These formulas can be extracted from \cite{Les1, \S5},
\cite{Les2, \S1.4}. Another way to get them is to use a general
method described in \cite{NU}. This method holds for any
orthogonal polynomials of hypergeometric type and allows to compute
all the data starting from the differential equation. In 
our case the differential equation has the form
$$
-(1+x^2)\p_m''+2(-\Im s+(\Re s+N-1)x)\p_m'+m(m+1-2\Re s-2N)\p_m=0. 
\tag 1.8
$$
\qed
\enddemo 

Note the symmetry property
$$
\p_m(-x)=(-1)^m\p_m(x)\mid_{s\leftrightarrow\bar s}\,. \tag1.9
$$
It follows from the symmetry of the weight function
$$
\phi(-x)=\phi(x)\mid_{s\leftrightarrow\bar s}
$$
and can be verified directly from the expression \tht{1.6}. 

To compute the Christoffel--Darboux kernel we will use the
classical formula 
$$
\sum_{m=0}^{N-1}\frac{\p_m(x')\p_m(x'')}{\Vert \p_m\Vert^2}=\frac 1{\Vert
\p_{N-1}\Vert^2}\,\frac{\p_N(x')\p_{N-1}(x'')-\p_{N-1}(x')\p_N(x'')}
{x'-x''}\,. 
\tag1.10
$$
If the parameter $s$ satisfies the stronger condition $\Re s>\frac 12$
then the polynomial $\p_N(x)$ exists and the formula holds. Since all
the terms in the left--hand side depend analytically on $s$ and
$\bar{s}$, we can use the formula for 
$s$ with $\frac 12\ge\Re s>-\frac 12$ as well with understanding that
the kernel is obtained by analytic continuation in $s$ and $\bar s$
viewed as independent variables (or, equivalently, by analytic
continuation in the variables $s$ and $s+\bar s$). 

Note that the trick with analytic continuation is actually needed
only for the values of $s$ on the vertical line $\Re s=0$, because
a singularity in the expression \tht{1.6} for $m=N$ arises for  
$\Re s=0$ only. 

The next lemma makes it possible to get an alternative expression for the
Christoffel--Darboux kernel. Its advantage is that all the terms have
no singularity in the whole region $\Re s>-\frac12$. 

\proclaim{Lemma 1.3} Set
$$
\wt\p_N(x)=\p_N(x)-\frac{2iNs}{2\Re s(2\Re s+1)}\p_{N-1}(x). \tag1.11
$$
This polynomial, initially defined for $\Re s>\frac12$, actually
makes sense for $\Re s>-\frac12$, as follows from the explicit formula 
$$
\wt\p_N(x)=(x-i)^N{}_2F_1\left[ \matrix -N+1,\, s+1\\ 
2\Re s+2\endmatrix\Biggr|\frac 2{1+ix}\right]. \tag1.12
$$
\endproclaim

\demo{Proof} Indeed, using the power series expansion of the
hypergeometric function it is readily verified that the following
general relation holds:
$$
{}_2F_1\left[ \matrix a,\, b\\ 
c\endmatrix\Biggr|\,z\,\right]=
{}_2F_1\left[ \matrix a,\, b\\ 
c+1\endmatrix\Biggr|\,z\,\right]+
\frac{abz}{c(c+1)}\,{}_2F_1\left[ \matrix a+1,\, b+1\\ 
c+2\endmatrix\Biggr|\,z\,\right]\,. \tag1.13
$$
{}From \tht{1.13} and \tht{1.6} we easily get \tht{1.12}. \qed
\enddemo

We summarize the above results in the following

\proclaim{Theorem 1.4} The correlation functions of the
$N$th pseudo--Jacobi ensemble \tht{1.2} have
the form 
$$
\r_n^{(s,N)}(x_1,\dots,x_n)=\det[\K^{(s,N)}(x_i,x_j)]_{i,j=1}^n \tag1.14
$$ 
with a kernel $\K^{(s,N)}(x',x'')$ defined on $\R\times\R$.

This kernel is given by the formulas 
$$
\multline
\K^{(s,N)}(x',x'')= \frac{2^{2\Re s}}{\pi}\,
\Gamma\left[\matrix 2\Re s+N+1,\, s+1,\,\bar s+1\\
N,\,2\Re s+1,\, 2\Re s +2\endmatrix\right]\\ \times
\frac{\p_N(x')\p_{N-1}(x'')-\p_{N-1}(x')\p_N(x'')}{x'-x''}\,
\sqrt{\phi(x')\phi(x'')}
\endmultline  \tag1.15
$$
or, equivalently,
$$
\multline
\K^{(s,N)}(x',x'')= \frac{2^{2\Re s}}{\pi}\,
\Gamma\left[\matrix 2\Re s+N+1,\, s+1,\,\bar s+1\\
N,\,2\Re s+1,\, 2\Re s +2\endmatrix\right]\\ \times
\frac{\wt\p_N(x')\p_{N-1}(x'')-\p_{N-1}(x')\wt\p_N(x'')}{x'-x''}\,
\sqrt{\phi(x')\phi(x'')}\,,
\endmultline \tag1.16
$$
where
$$
\phi(x)=(1+ix)^{-s-N}(1-ix)^{-\bar s-N}
=(1+x^2)^{-\Re s-N} e^{2\Im s\Arg(1+ix)}, \qquad x\in\R, \tag1.17
$$
and
$$
\gather
\p_{N}(x)=(x-i)^N\,{}_2F_1\left[\matrix -N,\, s\\ 2\Re s\endmatrix \, 
\Biggl|\,\frac 2{1+ix}\right], \tag1.18\\
\p_{N-1}(x)=(x-i)^{N-1}\,
{}_2F_1\left[\matrix -N+1,\, s+1\\ 2\Re s+2\endmatrix 
\, \Biggl|\,\frac 2{1+ix}\right], \tag1.19\\
\wt\p_N(x)=(x-i)^N{}_2F_1\left[ \matrix -N,\, s\\ 
2\Re s+2\endmatrix\Biggr|\frac 2{1+ix}\right]. \tag1.20
\endgather
$$
\endproclaim

Note that the expression \tht{1.15} is directly applicable when the parameter
$s$ does not lie on the line $\Re s=0$ while the expression
\tht{1.16} makes sense for any $s$ with $\Re s>-\frac12$. 

\demo{Proof} A standard argument from the Random Matrix Theory, see, e.g.,
\cite{Me} shows that the correlation functions are given by the
determinantal formula \tht{1.14}, where the kernel is equal to the
Christoffel--Darboux kernel \tht{1.5} multiplied by the factor
$\sqrt{\phi(x')\phi(x'')}$. Together with \tht{1.6}, \tht{1.7},
\tht{1.10} this implies the expression \tht{1.15} for the kernel. The
alternative formula \tht{1.16} then follows from Lemma 1.3. \qed 
\enddemo

\example{Remark 1.5} For $s=0$ the polynomial $\p_N$ can be defined
by taking the limit as $s\to0$ {\it along the real line.\/} {}From
the hypergeometric series it is easy to get the following expression: 
$$
\p_N(x)\mid_{s=0}=\frac{(x+i)^N+(x-i)^N}2\,.
$$
Likewise, we get
$$
\p_{N-1}(x)\mid_{s=0}=\frac{(x+i)^N-(x-i)^N}{2iN}\,.
$$
It follows that the Christoffel--Darboux kernel \tht{1.10} is an
elementary expression. This agrees with the fact that 
for $s=0$ our ensemble is related (via the Cayley transform) to the
Dyson ensemble.  
\endexample

\head 2. The scaling limit of the correlation functions \endhead

In this section we compute the scaling limit of the correlation functions 
of the pseudo--Jacobi ensemble as the number of particles goes to infinity.
The limit correlation functions have a determinantal form, and we express the
correlation kernel through the confluent hypergeometric function. 

Recall the definition of the confluent hypergeometric function:
$$
{}_1F_1\left[\matrix  a\\ c\endmatrix \, 
\Biggl|\,z\right]=\sum_{n=0}^\infty \frac{a(a+1)\dots(a+n-1)}
{c(c+1)\dots(c+n-1)\cdot n!}\,z^n\,,
$$
see, e.g., \cite{Er, 6.1}.

Let us rescale the correlation functions $\r_n^{(s,N)}$ of the
pseudo--Jacobi ensemble (see \tht{1.14}) by setting 
$$
\rho_n^{(s,N)}(x_1,\dots,x_n)=N^n\cdot\r_n^{(s,N)}(Nx_1,\dots,Nx_n).
$$
Note that the factor $N^n$ comes from the transformation of the
reference (Lebesgue) measure $dx_1\dots dx_n$. We will assume that
the variables range over the punctured real line $\R^*$, not the
whole line $\R$, as before.

\proclaim{Theorem 2.1} Let $\Re s>-\frac12$, as before. For any
$n=1,2,\dots$ and 
$x_1,\dots,x_n\in\R^*$ there exists a limit of the scaled $n$th correlation
functions $\rho_n^{(s,N)}$ as $N\to\infty$:
$$
\lim_{N\to\infty}\rho_n^{(s,N)}(x_1,\dots,x_n)
=\det\left[K^{(s,\infty)}(x_i,x_j)\right]_{1\le i,j\le n}.
$$
Here the kernel $K^{(s,\infty)}(x',x'')$ on $\R^*\times\R^*$ is as follows
$$
\gathered
K^{(s,\infty)}(x',x'')=\frac 1{2\pi}\,\Gamma\left[\matrix  s+1,\,\bar 
s+1 \\
2\Re s+1,\, 2\Re s +2\endmatrix\right]
\frac{P(x')Q(x'')-Q(x')P(x'')}{x'-x''}\,,
\\  
P(x)=\left|\frac 2x\right|^{\Re s}{e^{-i/x+\pi\Im s\cdot
\sgn(x)/2}}\,{}_1F_1\left[\matrix  s\\ 2\Re s\endmatrix \, 
\Biggl|\,\frac{2i}x\right],\\
Q(x)=\frac 2x\,\left|\frac 2x\right|^{\Re s}{e^{-i/x+\pi\Im s\cdot
\sgn(x)/2}}\,{}_1F_1\left[\matrix  s+1\\ 2\Re s+2\endmatrix \, 
\Biggl|\,\frac{2i}x\right].
\endgathered \tag2.1
$$
Or, equivalently,
$$
\gathered
K^{(s,\infty)}(x',x'')=\frac 1{2\pi}\,\Gamma\left[\matrix  s+1,\,\bar 
s+1 \\
2\Re s+1,\, 2\Re s +2\endmatrix\right]
\frac{\wt P(x')Q(x'')-Q(x')\wt P(x'')}{x'-x''}\,,
\\  
\wt P(x)=\left|\frac 2x\right|^{\Re s}{e^{-i/x+\pi\Im s\cdot
\sgn(x)/2}}\,{}_1F_1\left[\matrix  s\\ 2\Re s+1\endmatrix \, 
\Biggl|\,\frac{2i}x\right]
\endgathered \tag2.2
$$

The limit is uniform provided that the variables $x_1,\dots,x_n$ range
over any compact subset of $\R^*$.
\endproclaim

\demo{Comments} 1. As in Theorem 1.4, the first formula, \tht{2.1},
is directly applicable provided that $s$ does not lie on the line $\Re s=0$,
while the second formula, \tht{2.2}, holds for any $s$ with $\Re s>-\frac12$. 

2. The kernel $K^{(s,\infty)}(x',x'')$ can be
expressed through the M--Whittaker functions, see \cite{Er, 6.9} for
the definition. Namely, 
$$
P(x)=e^{-\frac{i\pi\bar{s}\,\sgn(x)}2}
M_{-i\Im s,\Re s-\frac 12}\left(\frac {2i}x\right),\quad 
Q(x)=e^{-\frac{i\pi(\bar{s}+1)\sgn(x)}2}
M_{-i\Im s,\Re s+\frac 12}\left(\frac {2i}x\right).
\tag 2.3
$$

3. The symmetry property \tht{1.9} of the pseudo--Jacobi polynomials
implies that
$$
P(-x)=P(x)\mid_{s\leftrightarrow \bar s}\,, \qquad 
Q(-x)=-Q(x)\mid_{s\leftrightarrow \bar s}\,, \tag2.4
$$
which can also be verified directly from \tht{2.3} by making use of
the formula \cite{Er, 6.9(7)}: 
$$
M_{\ka,\mu}(t)=e^{i\epsilon\pi\left( \mu+\frac 12\right)}
M_{-\ka,\mu}(-t), \quad \epsilon=\cases 1,&\Im t>0,\\-1,&\Im t<0.\endcases
$$
It follows that the correlation kernel
$K^{(s,\infty)}(x',x'')$ remains invariant when $x',x'',s$ are replaced by
$-x',-x'', \bar s$ (there is one more change of sign in the denominator
$(x'-x'')$).  

4. Formula \tht{2.4} implies that the functions $P(x)$ and $Q(x)$
are real--valued, which agrees with the fact that the pseudo--Jacobi
polynomials have real coefficients. Hence, the kernel
$K^{(s,\infty)}(x',x'')$ is real symmetric. 

5. When $s$ is real, the confluent hypergeometric function ${}_1F_1$
turns into the Bessel function, and the expressions for $P$ and $Q$
can be written as follows:
$$
\gather
P(x)=2^{2s-1/2}\Ga(s+1/2)|x|^{-1/2}J_{s-1/2}\left(\frac1{|x|}\right),\\
Q(x)=\sgn(x)2^{2s+1/2}\Ga(s+3/2)|x|^{-1/2}J_{s+1/2}\left(\frac1{|x|}\right).
\endgather
$$

6. For $s=0$ the Bessel functions with indices $\pm\frac12$
degenerate to trigonometric functions, and we get
$$
\gather
P(x)\mid_{s=0}=\cos(\tfrac1x), \qquad Q(x)\mid_{s=0}=2\sin(\tfrac1x), \\
K^{(0,\infty)}(x',x'')=\frac1\pi\,\frac{\sin(\frac1{x''}-\frac1{x'})}
{{x'}-{x''}}\,.
\endgather
$$
Changing the variable, $y=\frac1{\pi x}$, and taking into account the
corresponding transformation of the differential $dx$ we get the sine
kernel, in accordance with \tht{0.9}.  

\enddemo  

\demo{Proof of Theorem 2.1}
We will show that
$$
\lim_{N\to\infty} (\sgn(x')\sgn(x''))^{N}N\cdot K_s^{(N)}(Nx',Nx'')
=K_s^{(\infty)}(x',x''),\quad x',x''\in\R^*,
$$
uniformly on compact sets in $\R^*$. Note that the factor
$(\sgn(x')\sgn(x''))^N$ does not affect the determinantal formula. 

We start with the formula \tht{1.15}. First of all, we
remark that 
$$
\frac{\Gamma(2\Re s+N+1)}{\Gamma(N)}\sim N^{2\Re s+1},
$$
which easily follows from the Stirling formula. 

Next, we will examine the asymptotics of 
$$
\p_N(Nx)\sqrt{\phi(Nx)}, \quad  \p_{N-1}(Nx)\sqrt{\phi(Nx)},
\qquad N\to\infty.
$$
Here we will assume that $x$ is not a real but a complex variable
ranging in a neighborhood of a point $x_0\in\R^*$. This will allow us
to overcome the difficulty related to the singularity $x'-x''=0$ in the
denominator of \tht{1.15} by making use of the Cauchy formula.    

The asymptotics of the hypergeometric functions entering the formulas
\tht{1.18} and \tht{1.19} is as follows: 
$$
\gather
\lim_{N\to\infty}{}_2F_1\left[\matrix -N,\, s\\ 2\Re s\endmatrix \, 
\Biggl|\,\frac 2{1+iNx}\right]=
{}_1F_1\left[\matrix  s\\ 2\Re s\endmatrix\, 
\Biggl|\,\frac{2i}x\right], \\
\lim_{N\to\infty}{}_2F_1\left[\matrix -N+1,\, s+1\\ 2\Re s+2\endmatrix \, 
\Biggl|\,\frac 2{1+iNx}\right]=
{}_1F_1\left[\matrix  s+1\\ 2\Re s+2\endmatrix\, 
\Biggl|\,\frac{2i}x\right].
\endgather
$$
Indeed, this is a special case of the well--known limit relation
$$
\lim_{|a|\to\infty}{}_2F_1\left[\matrix a,\, b\\ c\endmatrix \, 
\Biggl|\,\frac za\right]=
{}_1F_1\left[\matrix  b\\ c\endmatrix\, 
\Biggl|\,z\right], \qquad z\in\C.
$$
This can be readily verified by making use of the integral
representation of the hypergeometric function written in the form
$$
{}_2F_1\left[\matrix a,\, b\\ c\endmatrix \, 
\Biggl|\,\frac za\right]
=\Ga(c)\left\langle
\frac{t^{b-1}_+}{\Ga(b)}\frac{(1-t)^{c-b-1}_+}{\Ga(c-b)}\,, 
\frac1{(1-tz/a)^a}\right\rangle,
$$
where the brackets denote the pairing between a generalized function
(which in the present case is supported by $[0,1]$) and a test
function, and $t$ is the argument of the both functions. 
Note that the limit is uniform provided that $z$ ranges over a bounded
set of $\C$.  

The asymptotics of the remaining terms looks as follows: 
$$
\lim_{N\to\infty}(\pm1)^N(Nx-i)^N\sqrt{\phi(Nx)}\sim
N^{-\Re s}(\pm x)^{-\Re s}e^{-i/x}e^{\pm\pi\Im s}\,,
$$
where $\pm$ is the sign of $\Re x$ and the limit is uniform on
compact subsets in the open right or left half--plane. 
Indeed, assume $\Re x>0$. In the transformations below
any expression of the form $z^c$ with $c\in\C$ is understood as a
holomorphic function in the domain $\C\setminus(-\infty,0]$. We have 
$$
(Nx-i)^N\sqrt{\phi(Nx)}
=(Nx-i)^N(1+iNx)^{-(s+N)/2}(1-iNx)^{-(\bar s+N)/2}
$$
$$
\multline
=(Nx)^N (iNx)^{-(s+N)/2}(-iNx)^{-(\bar s+N)/2}\\
\times\left(1-\frac i{Nx}\right)^N
\left(1+\frac1{iNx}\right)^{-(s+N)/2}
\left(1-\frac1{iNx}\right)^{-(\bar s+N)/2}
\endmultline
$$
$$
\multline
=N^{-\Re s}x^{-\Re s}i^{-(s+N)/2}(-i)^{-(\bar s+N)/2}\\
\times\left(1-\frac i{Nx}\right)^N
\left(1+\frac1{iNx}\right)^{-(s+N)/2}
\left(1-\frac1{iNx}\right)^{-(\bar s+N)/2} 
\endmultline
$$
$$
\sim N^{-\Re s}x^{-\Re s}
e^{\pi\Im s}e^{-i/x}.
$$
For $\Re x<0$ the argument is similar.

Combining all these asymptotics we get the desired result. \qed
\enddemo

\head 3. The Hua--Pickrell measures \endhead

In this section we define the Hua--Pickrell measures. They form
a 2--parameter family of $U(\infty)$--invariant probability measures
on the space of infinite Hermitian matrices. 

Let $H(N)$ denote the real vector space formed by complex Hermitian
$N\times N$ matrices, $N=1,2,\dots$. 
Let $H$ stand for the space of all infinite Hermitian matrices
$X=[X_{i,j}]_{i,j=1}^\infty$. For $X\in H$ and $N=1,2,\dots$, we
denote by $\theta_N(X)\in H(N)$ the upper left $N\times N$ corner of
$X$. Using the projections $\theta_N\:H\to H(N)$, $N=1,2,\dots$, we
may identify $H$ with the projective limit space $\varprojlim H(N)$.
We equip $H$ with the corresponding projective limit topology. We
will also use the Borel structure on $H$ generated by this topology. 

Let $U(N)$ be the group of unitary $N\times N$ matrices,
$N=1,2,\dots$. For any $N$, we embed $U(N)$ into $U(N+1)$ using the
mapping $u\mapsto\bmatrix u&0\\ 0&1\endbmatrix$. Let
$U(\infty)=\varinjlim U(N)$ denote the corresponding inductive limit
group. We regard $U(\infty)$ as the group of infinite unitary
matrices $U=[U_{ij}]_{i,j=1}^{\infty}$ with finitely many
entries $U_{ij}\ne\delta_{ij}$. The group $U(\infty)$ acts on the
space $H$ by conjugations.

\proclaim{Proposition 3.1} For any $s\in\C$, $\Re s>-\tfrac12$, there
exists a probability Borel measure $\ms$ on $H$, characterized by the
following property: for any $N=1,2,\dots$, the image of $\ms$ under
the projection $\th_N$ is the probability measure $\msN$ on $H(N)$
defined by 
$$
\gathered
\msN(dX)=(\const_N)^{-1}\det((1+iX)^{-s-N})\det((1-iX)^{-\bar s-N})\\
\times \prod_{j=1}^N dX_{jj}
\prod_{1\le j<k\le N}d(\Re X_{jk})d(\Im X_{jk}),\\
\text{where}\quad \const_N=\prod_{j=1}^N
\frac{\pi^j\Ga(s+\bar s+j)}
{2^{s+\bar s+2j-2}\Ga(s+j)\Ga(\bar s+j)}\,. 
\endgathered
\tag3.1
$$

The measure $\ms$ is invariant under the action of $U(\infty)$.
\endproclaim

\demo{Comments} 1. For $X\in H(N)$ and $z\in\C$ we define the matrix
$(1\pm iX)^z$ by means of the functional calculus. This makes the expression
$$
f_N(X)=\det((1+iX)^{-s-N})\det((1-iX)^{-\bar s-N}), \qquad X\in H(N)
$$
meaningful. Equivalently, denoting by $x_1,\dots,x_N$ the eigenvalues
of $X$, 
$$
f_N(X)=\prod_{j=1}^N (1+ix_j)^{-s-N}(1-ix_j)^{-\bar s-N}\,, \tag3.2
$$
where we use the analytic continuation of the function $(\dots)^z$
from the positive axis to the region $\C\setminus(-\infty,0]$.  

2. When $s$ is real, the expression \tht{3.2} takes a simpler form
$$
f_N(X)=(\det(1+X^2))^{-s-N}\,, \qquad X\in H(N),\quad s\in\R. 
$$
\enddemo

\demo{Proof} Step 1. First of all, note that $f_N(X)\ge0$. Therefore, if
$f_N$ is integrable then it defines a finite measure on $H(N)$. 

Fix $N\ge2$ and write an arbitrary matrix $X\in H(N)$ in the block
form 
$$
X=\bmatrix Y & \xi \\ \xi^* & t\endbmatrix,
\qquad Y\in H(N-1), \quad \xi\in\C^{N-1}, \quad t\in\R. 
$$
We shall prove that for any $Y\in H(N-1)$ the integral of $f_N$ over
$\xi,t$ is finite and has the following value
$$
\multline
\int_{(\xi,t)\in\C^{N-1}\times\R}
f_N\left(\bmatrix Y & \xi \\ \xi^* & t\endbmatrix\right)\cdot 
\prod_{j=1}^N d(\Re\xi_j)d(\Im\xi_j) \cdot dt\\
=f_{N-1}(Y)\cdot \frac{\pi^N\Ga(s+\bar s+N)}
{2^{s+\bar s+2N-2}\Ga(s+N)\Ga(\bar s+N)}\,. 
\endmultline\tag3.3
$$

For $N=1$, $Y$ and $t$ disappear, and the claim is that the integral
of $f_1$ over $\R$ is finite and given by
$$
\int_{t\in\R} f_1(t)dt
=\int_{-\infty}^\infty  (1+it)^{-s-1}(1-it)^{-\bar s-1}dt
=\frac{\pi \Ga(s+\bar s+1)}{2^{s+\bar s}\Ga(s+1)\Ga(\bar s+1)}\,.
\tag3.4
$$

Let us show that \tht{3.3} and \tht{3.4} imply the proposition.
Indeed, using induction on $N$ we see that the integral of $f_N$ over
$H(N)$ is finite and equals $\const_N$. Thus, the measure $\msN$ is
correctly defined for any $N$.

Next, \tht{3.3} implies that the measures $\msN$ and $m^{(s,N-1)}$
are consistent with the projection $X\mapsto Y$ from $H(N)$ to
$H(N-1)$. Since $H$ coincides with the projective limit of
the spaces $H(N)$ as $N\to\infty$, we conclude that the measure $\ms$
exists and is unique.

Finally, $\ms$ is invariant under the action of $U(\infty)$, because
each $\msN$ is invariant under the action of $U(N)$ for all $N=1,2,\dots$. 

Step 2. We proceed to the proof of \tht{3.3} and \tht{3.4}. The
latter formula follows from formula \tht{3.9} in Lemma 3.3. The
former formula is proved in \cite{Hua, Theorem 2.1.5} for real $s$, 
and we employ his argument with slight
modifications. Applying Lemma 3.2 (see below) we get 
$$
\multline
f_N(X)=\det((1+iY)^{-s-N})(1+it+\xi^*(1+iY)^{-1}\xi)^{-s-N}\\
\times\det((1-iY)^{-\bar s-N})(1-it+\xi^*(1-iY)^{-1}\xi)^{-\bar s-N}\,. 
\endmultline
\tag3.5
$$

Next, note that the integral \tht{3.3} is invariant under the conjugation
of $Y$ by a matrix $V\in U(N-1)$. Indeed to see this, we use the
invariance of the function $f_N$ and make a change of a variable,
$V\xi\mapsto\xi$. Therefore, without loss of generality we may assume
that $Y$ is a diagonal matrix. Denoting its diagonal entries (which
are real numbers) as $y_1,\dots,y_{N-1}$ and using \tht{3.5} we
reduce the integral \tht{3.3} to
$$
\multline
\prod_{j=1}^{N-1}(1+iy_j)^{-s-N}(1-iy_j)^{-\bar s-N}\\
\times
\int_{(\xi,t)\in\C^{N-1}\times\R}
\left(1+\sum_{j=1}^{N-1}\frac{|\xi_j|^2}{1+y_j^2}
+i\left(t-\sum_{j=1}^{N-1}\frac{|\xi_j|^2 y_j}{1+y_j^2}\right)
\right)^{-s-N}\\
\times \left(1+\sum_{j=1}^{N-1}\frac{|\xi_j|^2}{1+y_j^2}
-i\left(t-\sum_{j=1}^{N-1}\frac{|\xi_j|^2 y_j}{1+y_j^2}\right)
\right)^{-\bar s-N}
\prod_{j=1}^N d(\Re\xi_j)d(\Im\xi_j) \cdot dt.
\endmultline\tag3.6
$$

This integral is easily simplified. First, assuming the variables
$\xi_1,\dots,\xi_{N-1}$ fixed, we make a change of variable
$$
t-\sum_{j=1}^{N-1}\frac{|\xi_j|^2 y_j}{1+y_j^2}\,\mapsto\,t. 
$$
Next, we change the variables $\xi_j$,
$$
\frac{\xi_j}{\sqrt{1+y_j^2}}\, \mapsto\,\xi_j\,, \qquad
j=1,\dots,N-1,  
$$
which gives rise to the factor $\prod(1+y_j)^2$. Then \tht{3.6} is
reduced to 
$$
\multline
\prod_{j=1}^{N-1}(1+iy_j)^{-s-N+1}(1-iy_j)^{-\bar s-N+1}\cdot
\int_{(\xi,t)\in\C^{N-1}\times\R}
\left(1+\sum_{j=1}^{N-1}|\xi_j|^2+it\right)^{-s-N}\\
\times\left(1+\sum_{j=1}^{N-1}|\xi_j|^2-it\right)^{-\bar s-N}
\prod_{j=1}^N d(\Re\xi_j)d(\Im\xi_j) \cdot dt.  
\endmultline \tag3.7
$$
Setting $r=\sum|\xi_j|^2$ we readily reduce \tht{3.7} to
$$
\multline
\prod_{j=1}^{N-1}(1+iy_j)^{-s-N+1}(1-iy_j)^{-\bar s-N+1}\\
\cdot \frac{\pi^{N-1}}{\Ga(N-1)}\,
\int_{r\ge0}\int_{t\in\R}(1+r+it)^{-s-N}(1+r-it)^{-\bar s-N}r^{N-2}drdt.
\endmultline
$$
By Lemma 3.3, the double integral is finite and its value is given by
formula \tht{3.9} where we substitute $a=s+N$, $b=\bar s+N$ (the
assumption of Lemma 3.3 is satisfied because $\Re s>-\tfrac12$). This
implies \tht{3.3}.  
\qed
\enddemo

We proceed to the proof of two lemmas which were used in Proposition
3.1. 

\proclaim{Lemma 3.2} Consider the $N\times N$ matrix analog of the
right halfplane in $\C$:
$$
\Mplus=\{A\in\M\mid A+A^*>0\}.
$$
Write $N\times N$ matrices in the block form according to a partition
$N=N_1+N_2$,
$$
A=\bmatrix A_{11} & A_{12}\\ A_{21} & A_{22}\endbmatrix\,. 
$$
Then for $z\in\C$ and $A\in\Mplus$ the following relation holds
$$
\det(A^z)=\det(A_{11}^z)\det((A_{22}-A_{21}A_{11}^{-1}A_{12})^z).
\tag3.8
$$
\endproclaim

\demo{Proof} First of all, we show that both sides in \tht{3.8} make
sense. Note that if $A\in\Mplus$ then any eigenvalue $\la$ of $A$
lies in the open right halfplane (indeed, if $\xi\in\C^N$ is an
eigenvector with the eigenvalue $\la$ then
$0<((A+A^*)\xi,\xi)=2\Re\la(\xi,\xi)$, which implies $\Re\la>0$).
Therefore, we can define the matrix $A^z$ by means of the functional
calculus. Next, note that the matrices $A_{11}$ and 
$A_{22}-A_{21}A_{11}^{-1}A_{12}$ also belong to the matrix right
halfplanes. Indeed, for the former matrix this is evident, and for
the latter matrix this follows from the fact that $A^{-1}\in\Mplus$
and
$$
A_{22}-A_{21}A_{11}^{-1}A_{12}=((A^{-1})_{22})^{-1}\,. 
$$
Thus, the expressions $(\dots)^z$ in the right--hand side of
\tht{3.8} are well--defined.

Since both sides of \tht{3.8} are holomorphic functions in $A$ in
the connected region $\Mplus$, we may assume, without loss of
generality, that $A$ lies in a small neighborhood of the matrix 1.
Then we may interchange the symbol of determinant and exponentiation.
This reduces \tht{3.8} to the classical formula for the determinant
of a block matrix,
$$
\det A=\det A_{11}\cdot \det(A_{22}-A_{21}A_{11}^{-1}A_{12}).
$$
See, e.g. \cite{Gantmakher, Ch. II, \S5.3}.
\qed
\enddemo

\proclaim{Lemma 3.3} We have
$$
\multline
\frac{\pi^{N-1}}{\Ga(N-1)}\,
\int_{r\ge0}\int_{t\in\R}(1+r+it)^{-a}(1+r-it)^{-b}r^{N-2}drdt\\
=\frac{\pi^N\Ga(a+b-N)}{2^{a+b-2}\Ga(a)\Ga(b)}\,, 
\qquad a,b\in\C, \quad \Re(a+b)>N, \quad N>1.
\endmultline \tag3.9
$$
and
$$
\int_{t\in\R}(1+it)^{-a}(1-it)^{-b}dt
=\frac{\pi\Ga(a+b-1)}{2^{a+b-2}\Ga(a)\Ga(b)}\,, \qquad
a,b\in\C, \quad \Re(a+b)>1. \tag3.10
$$
\endproclaim

\demo{Proof} The integral \tht{3.10} is readily reduced to a known
integral, see \cite{Er, 1.5 (30)}. 

To evaluate the integral
\tht{3.9}, make a change of variable, $t\mapsto(1+r)t$. The
integral splits into the product of two integrals, one of which is
\tht{3.10} and the other one is the integral
$$
\int_{r\ge0}(1+r)^{-a-b+1}\frac{r^{N-2}}{\Ga(N-1)}dr
=\frac{\Ga(a+b-N)}{\Ga(a+b-1)}\,. 
$$
This proves \tht{3.9}.

Note also that \tht{3.10} is a degeneration of \tht{3.9}, because
$r_+^{N-2}/\Ga(N-1)$ degenerates to the delta function $\de(r)$ at
$N=1$.  
\qed
\enddemo

Let $\C_+$ denote the right halfplane. Following Neretin
\cite{Ner2} we define a map 
$$
H\ni X=[X_{jk}]_{j,k=1}^\infty
\mapsto (\zeta_1,\zeta_2,\dots)\in\R\times\C_+^\infty \tag3.11
$$ 
as follows. For any $N=2,3,\dots$, write the matrix
$\th_N(X)=[X_{jk}]_{j,k=1}^N$ in the block form
$$
\th_N(X)=\bmatrix \th_{N-1}(X) & \xi\\ \xi^* & t\endbmatrix 
$$
and then set 
$$
\zeta_N=it+\xi^*(1+i\th_{N-1})^{-1}\xi\in\C_+. 
$$
Finally, set $\zeta_1=X_{11}\in\R$.  

\proclaim{Proposition 3.4} The pushforward of the measure $\ms$ under
the map \tht{3.11} is a product measure $\mu_1\times\mu_2\times\dots$
on the space $\R\times\C_+^\infty$. Here $\mu_1, \mu_2,\dots$ are the
following probability measures:
$$
\mu_1(dt)=\frac{2^{s+\bar s}\Ga(s+1)\Ga(\bar s+1)}
{\pi\Ga(s+\bar s+1)}(1+it)^{-s-1}(1-it)^{-\bar s-1}dt 
$$
and, for $N\ge2$, $\zeta=r+it\in\C_+$,
$$
\mu_N(d\zeta)=\frac{2^{s+\bar s+2N-2}\Ga(s+N)\Ga(\bar s+N)}
{\pi\Ga(s+\bar s+N)}\,(1+\zeta)^{-s-N}(1+\bar \ze)^{-\bar s-N}
\frac{r^{N-2}}{\Ga(N-1)}drdt. \tag3.12
$$
\endproclaim

\demo{Proof} This follows from the proof of Proposition 3.1. \qed
\enddemo

\proclaim{Theorem 3.5} The Hua--Pickrell measures $\ms$ are pairwise
disjoint. I.e., if $s'$, $s''$ are two distinct values of the parameter $s$
then there exist two disjoint Borel sets in $H$ supporting the
measures $m^{(s')}$ and $m^{(s'')}$, respectively.
\endproclaim

\demo{Proof} We will apply Kakutani's theorem \cite{Ka}. Assume
first we are given two probability measures, $\mu'$ and $\mu''$,
defined on the same Borel space. Take any measure $\nu$ such that
both $\mu'$ and $\mu''$ are absolutely continuous with respect to
$\nu$. For instance, $\nu=\mu'+\mu''$. Denote by $\mu'/\nu$ and
$\mu''/\nu$ the respective Radon--Nikodym derivatives. The measure 
$\sqrt{\frac{\mu'}\nu\,\frac{\mu''}\nu}\cdot\nu$ does not depend on the
choice of $\nu$. Denote it by $\sqrt{\mu'\mu''}$ and set
$$
\langle \mu',\mu''\rangle=\int\sqrt{\mu'\mu''}.
$$
We have $0\le \langle \mu',\mu''\rangle\le1$. Moreover, 
$\langle\mu',\mu''\rangle=1$ is equivalent to $\mu'=\mu''$ while 
$\langle\mu',\mu''\rangle=0$ exactly means that $\mu'$ and $\mu''$
are disjoint. 

Next, assume 
$\mu'=\mu'_1\times\mu'_2\times\dots$ and 
$\mu''=\mu''_1\times\mu''_2\times\dots$ are two product
probability measures defined on the same countably infinite
product space. Kakutani's theorem \cite{Ka} says that $\mu'$ and
$\mu''$ are disjoint if the infinite product
$\prod_{N=1}^\infty \langle \mu'_N,\mu''_N\rangle$
is divergent, i.e., the partial products tend to 0. 

Finally, consider the product space $\R\times\C_+^\infty$ and take as
$\mu'$ and $\mu''$ the pushforwards of measures $m^{(s')}$ and $m^{(s'')}$,
respectively, as explained in Proposition 3.4. We prove that
$\mu'$ and $\mu''$ are disjoint. Then this immediately implies the
same for the initial measures $m^{(s')}$ and $m^{(s'')}$.

We drop the index $N=1$ which plays a special role and calculate the
integral defining $\langle \mu'_N,\mu''_N\rangle$ for $N\ge2$. By \tht{3.12}
and \tht{3.9} we get $$
\gather
\langle\mu'_N,\mu''_N\rangle
=\sqrt{\frac{\Ga(s'+N)\Ga(\overline{s'}+N)\Ga(s''+N)\Ga(\overline{s''}+N)}
{\Ga(s'+\overline{s'}+N)\Ga(s''+\overline{s''}+N)}}\,
\frac{\Ga(s+\bar s+N)}{\Ga(s+N)\Ga(\bar s+N)}\,, \\
s=\frac{s'+s''}2\,. 
\endgather
$$
The classical asymptotic formula for the ratio of two
$\Ga$-functions, see \cite{Er, 1.18(4)}, implies that 
$$
\frac{\Ga(z+N)\Ga(\bar z+N)}{\Ga(z+\bar z+N)\Ga(N)}\sim
1-\frac{z\bar z}N+O\left(\frac1{N^2}\right)\,. 
$$
It follows that
$$
\langle\mu'_N,\mu''_N\rangle
\sim 1-\frac{|s'-s''|^2}{4N}+O\left(\frac1{N^2}\right). 
$$
Thus, the product of $\langle\mu'_N,\mu''_N\rangle$'s is divergent. \qed
\enddemo

\head 4. Ergodic measures \endhead

In this section we recall the classification theorem and some other known
results on $U(\infty)$--invariant ergodic probability measures on the space of
infinite Hermitian matrices. 

Consider the natural embeddings
$$
H(N)\to H(N+1),\qquad A\mapsto\bmatrix A&0\\ 0&0\endbmatrix,
$$
and denote by $H(\infty)$ the corresponding inductive limit space
$\varinjlim H(N)$. Then $H(\infty)$ is identified with the space of infinite
Hermitian matrices with finitely many nonzero entries. We
equip $H(\infty)$ with the inductive limit topology. In particular, a
function $f\:H(\infty)\to{\Bbb C}$ is continuous if its restriction to
$H(N)$ is continuous for any $N$.

There is a natural pairing
$$
H(\infty)\times H\to\R,\qquad(A,X)\mapsto\tr(AX).
$$
$H$ is the algebraic dual space of $H(\infty)$ with respect to this
pairing. 

Using the map
$$
H\ni X\mapsto \{X_{ii}\}_{i=1}^\infty \sqcup 
\{\Re X_{ij}, \Im X_{ij}\}_{i<j}
$$
we can identify $H$, as a topological vector space, with the infinite
product space ${\Bbb R}^{\infty}={\Bbb R}\times{\Bbb R}\times\cdots$.
Under this identification, $H(\infty)\subset H$ turns into 
${\Bbb R}^{\infty}_0:=\bigcup_{n\ge1}{\Bbb R}^n$, and the pairing
defined above becomes the standard pairing between 
${\Bbb R}^{\infty}_0$ and ${\Bbb R}^{\infty}$.

Given a Borel probability measure $M$ on $H$, we define its Fourier
transform, or characteristic function, as the following function on
$H(\infty)$:
$$
A\mapsto\int_H e^{i\tr(AX)}M(dX).\tag4.1
$$

The group $U(\infty)$ acts by
conjugations both on $H(\infty)$ and $H$, and the pairing between
these two spaces is clearly $U(\infty)$-invariant.
Each matrix from $H(\infty)$ is conjugated to a diagonal matrix 
$\diag(r_1,r_2,\dots)$ with finitely many nonzero entries. It follows
that the Fourier transform of a $U(\infty)$-invariant measure on $H$
is uniquely determined by its values on diagonal matrices from
$H(\infty)$. 

Set 
$$
\gather
\Om=\{\om=(\al^+,\al^-,\ga_1,\de)\in\R^{2\infty+2}=
\R^\infty\times\R^\infty\times\R\times\R\mid\\
\al^+=(\al^+_1\ge \al^+_2\ge\dots\ge0),\qquad
\al^-=(\al^-_1\ge \al^-_2\ge\dots\ge0),\\
\ga_1\in\R,\qquad \de\ge0,\qquad 
\sum (\al^+_i)^2+\sum(\al^-_i)^2\le\de \}.
\endgather
$$
This is a closed region in $\R^{2\infty+2}$.

Denote
$$
\ga_2=\de-\sum (\al^+_i)^2-\sum (\al^-_i)^2\ge0.
$$
In this notation we have

\proclaim{Proposition 4.1} There exists a parametrization of
ergodic $U(\infty)$-invariant probability measures on the space $H$
by points of the space $\Om$. Given $\om$, the Fourier
transform \tht{4.1} of the corresponding ergodic measure $M^\om$ is
given by 
$$
\multline
\int_{X\in H}e^{i\,\tr(\diag(r_1,\dots,r_n,0,0,\dots)\,X)}M^\om(dX)\\
=\prod_{j=1}^n\left\{
e^{i\ga_1 r_j-\ga_2 r_j^2}
\prod_{k=1}^\infty \frac{e^{-i\al^+_k r_j}}{1-i\al^+_k r_j}\,
\prod_{k=1}^\infty \frac{e^{i\al^-_k r_j}}{1+i\al^-_k r_j}\right\}
\endmultline
$$
\endproclaim

\demo{Proof} See \cite{Pi2, Proposition 5.9} and \cite{OV, Theorem
2.9}. \qed 
\enddemo

\example{Remark 4.2}
If only one of the parameters $\al^\pm_i$, $\ga_1$, $\ga_2$ is
distinct from 0 then the 
corresponding ergodic measure is called {\it elementary.\/} See
\cite{OV, Corollaries 2.5--2.7} for a description of the elementary
measures. Note, in particular, that the elementary measures
corresponding to the parameter $\ga_2$ are standard Gaussian measures
on $H$, see \cite{OV, Corollary 2.6}. Since the
expression of Proposition 4.1 is multiplicative with respect to the
coordinates of $\om$, any ergodic measure is a convolution product
of elementary ergodic measures.  
\endexample

For $N=1,2,\dots$, let $\SS_N\subset\R^N$ denote the set of
$N$-tuples of weakly decreasing real numbers:
$$
\la=(\la_1\ge\dots\ge\la_N).
$$ 
Given $\la\in\SS_N$, let $\Orbb(\la)$ denote the set of matrices $X\in
H(N)$ with eigenvalues $\la_1,\dots,\la_N$. The sets of the form
$\Orbb(\la)$ are exactly the $U(N)$-orbits in $H(N)$. 

Given $\la\i\SS_N$, we set
$$
\gather
a^+_i(\la)=\cases \dfrac{\max(\la_i,0)}N, & i=1,\dots,N, \\
0, & i=N+1,N+2,\dots\,, \endcases\\
a^-_i(\la)=\cases \dfrac{\max(-\la_{N+1-i},0)}N, & i=1,\dots,N, \\
0, & i=N+1,N+2,\dots.\, \endcases
\endgather
$$
Equivalently, if $k$ and $l$ denote the numbers of strictly positive
terms in $\{a^+_i\}$ and $\{a^-_i\}$, respectively then
$$
\la=(a^+_1(\la),\dots,a^+_k(\la),0,\dots,0,-a^-_l(\la),\dots,-a^-_1(\la)).
$$

Further, we set
$$
\gather
c(\la)=\sum_{i=1}^\infty a^+_i(\la)-\sum_{i=1}^\infty a^-_i(\la)=
\frac{\la_1+\dots+\la_N}N\,,\\
d(\la)=\sum_{i=1}^\infty (a^+_i(\la))^2
+\sum_{i=1}^\infty(a^-_i(\la))^2 =
\frac{\la_1^2+\dots+\la_N^2}{N^2}\,.
\endgather
$$

By virtue of \cite{OV, Theorem 3.3}, any ergodic measure can be approximated by
orbital measures on the spaces $H(N)$ as $N\to\infty$. The next
result provides an explicit description of the approximating orbital
measures. It also clarifies the meaning of the parameters
in Proposition 4.1. 

\proclaim{Proposition 4.3} Let $\{\Orbb(\la^{(N)})\mid \la^{(N)}\in\SS_N\}$
be a sequence of orbits and let 
$\{M^{(N)}\}$ be the sequence of the corresponding orbital
measures on the spaces $H(N)$, $N=1,2,\dots$. We view each $M^{(N)}$ as a
measure on $H$.  

The measures $M^{(N)}$ weakly converge to a measure $M$ on $H$, i.e., 
$\langle f, M^{(N)}\rangle\to \langle f,M\rangle$ for any bounded
continuous function $f$ on $H$, if and only if there exist limits
$$
\gather
\al^\pm_i=\lim_{N\to\infty}a^\pm_i(\la^{(N)}), \qquad i=1,2,\dots, \\
\ga_1=\lim_{N\to\infty} c(\la^{(N)}), \\
\de=\lim_{N\to\infty} d(\la^{(N)}).
\endgather
$$

If this condition holds then the collection
$\om=(\{\al^+_i\},\{\al^-_i\},\ga_1,\de)$ is a point of $\Om$ and the limit
measure $M$ coincides with the ergodic measure $M^\om$.
\endproclaim

\demo{Proof} See \cite{OV, Theorem 4.1.} \qed
\enddemo

\proclaim{Proposition 4.4} For any $U(\infty)$-invariant probability measure
$M$ on $H$ there exists a probability measure $P$ on $\Om$ such that
$$
M=\int_\Om M^\om P(d\om),
$$
which means that for any bounded Borel function $f$ on $H$,
$$
\langle f,M\rangle=\int_\Om \langle f,M^\om\rangle P(d\om).
\tag 4.2
$$
Such measure $P$ is unique. Conversely, any probability measure $P$
on $\Om$ arises in this way from a certain measure $M$. 
\endproclaim

\demo{Proof} This follows from Theorem 9.1 and Proposition 9.4. \qed
\enddemo

We will call $P$ the {\it spectral measure\/} for $M$.

\head 5. Approximation of spectral measures \endhead

In this section we show that the spectral measure for a $U(\infty)$--invariant
probability measure $M$ on $H$ can be obtained as a certain limit of
finite--dimensional projections of $M$.  

For $X\in H$, let $\la^{(N)}(X)\in\SS_N$ denote the spectrum of the
finite matrix $\theta_N(X)\in H(N)$. Let us say that $X\in H$ is {\it
regular\/} if there exist limits
$$
\gathered
\al^\pm_i(X)=\lim_{N\to\infty}a^\pm_i(\la^{(N)}(X)), \qquad i=1,2,\dots, \\
\ga_1(X)=\lim_{N\to\infty} c(\la^{(N)}(X)), \\
\de(X)=\lim_{N\to\infty} d(\la^{(N)}(X)).
\endgathered \tag5.1
$$
Let $\Hreg\subset H$ denote the subset of regular matrices in $H$.
Since $\la^{(N)}(X)$ is a continuous function in $X$ for any $N$, the
functions $a^\pm_i(\la^{(N)}(X))$, $c(\la^{(N)}(X))$, and
$d(\la^{(N)}(X))$ are also continuous. It follows that $\Hreg$ is a
Borel subset of $H$ (more precisely, a subset of type
$F_{\sigma\de}$). 

\proclaim{Theorem 5.1} Any $U(\infty)$-invariant probability measure
on $H$ is supported by $\Hreg$. 
\endproclaim

\demo{Proof} First, let $M$ be an ergodic $U(\infty)$-invariant
probability measure on $H$. By Vershik's ergodic theorem (see \cite{OV, Theorem 3.2}), $M$ is concentrated
on the set of those $X\in H$ for which the orbital measures
$\Orbb(\la^{(N)}(X))$ weakly converge to $M$.
By Proposition 4.3, this set consists exactly of those $X$ for which the
limits \tht{5.1} exist and coincide with the parameters of $M$ given
in Proposition 4.1. All such matrices $X$ belong
to $\Hreg$, so that $M$ is supported by $\Hreg$. Thus, the claim of
the theorem holds for ergodic measures. 

Now let $M$ be an arbitrary $U(\infty)$-invariant probability measure on
$H$ and $P$ be its spectral measure. Apply \tht{4.2} by taking as
$f$ the characteristic function of the set $\Hreg\subset H$. We have
$\langle f,M^\om\rangle=1$ for any $\om\in\Om$. Since $P$ is a
probability measure, we get from \tht{4.2} that $\langle
f,M\rangle=1$. Therefore, $\Hreg$ is of full measure with respect to
$M$. \qed
\enddemo

Let $\pi:\Hreg\to\Om$ denote the map sending $X\in\Hreg$ to the point
$\om$ with the coordinates defined by \tht{5.1}. This is a Borel map,
because it is the pointwise limit of a sequence of continuous maps. 

\proclaim{Theorem 5.2} Let $M$ be a $U(\infty)$-invariant probability
measure on $H$ and let $M\mid_{\Hreg}$ be the restriction of $M$ to
$\Hreg$, which is correctly defined by Theorem 5.1. 

The pushforward of the measure $M\mid_{\Hreg}$ under the Borel map
$\pi$ introduced above coincides with the spectral measure $P$.
\endproclaim

\demo{Proof} Let $F$ be an arbitrary bounded Borel function on $\Om$
and $f$ be its pullback on $\Hreg$. We must prove that $\langle
f,M\rangle=\langle F,P\rangle$.

By definition of $P$, we have
$$
\langle f,M\rangle=\int_\Om \langle f,M^\om\rangle P(d\om).
$$
On the other hand, we know that for any
$\om\in\Om$, the measure $M^\om$ is supported by
$\pi^{-1}(\om)\subset\Hreg$ (see the beginning of the proof of
Theorem 5.1). Finally, by the definition of $f$, we
have $f\mid_{\pi^{-1}(\om)}\equiv F(\om)$, so that 
$\langle f,M^\om\rangle=F(\om)$.  

Therefore, the integral in the right--hand side is equal to 
$\langle F,P\rangle$. \qed
\enddemo

For $N=1,2,\dots$, let $\pi_N:H\to\Om\subset\R^{2\infty+2}$ denote
the composition of the maps $H\ni X\mapsto\la^{(N)}(X)\in\SS_N$ and 
$\SS_N\ni\la\mapsto(\{a^+_i(\la)\},\{a^-_i(\la)\},c(\la),d(\la))\in\Om$. 

\proclaim{Theorem 5.3} Let $M$ be a $U(\infty)$-invariant probability
measure on $H$, $P$ be its spectral measure, and $P_N$ be the
pushforward of $M$ under the map $\pi_N:H\to\Om$ defined above.

Then $P_N$ weakly converge to $P$ as $N\to\infty$. That is, for any
continuous bounded function $F$ on $\Om$, 
$$
\lim_{N\to\infty}\langle F,P_N\rangle \to \langle F,P_N\rangle.
$$
\endproclaim

\demo{Proof} By Theorem 5.1, $\Hreg\subset H$ is of full measure
with respect to $M$, so that we may view $(\Hreg,M)$ as a probability space.

We have
$$
\pi_N(M)=P_N, \qquad \pi(M)=P.
$$

Indeed, the first equality follows from the definition of $P_N$ and
the fact that $\Hreg$ is of full measure, and the second equality is
given by Theorem 5.2.

Next, by the very definition of $\Hreg$, we have $\pi_N(t)\to\pi(t)$
for any $t\in\Hreg$ as $N\to\infty$, where the limit is taken with
respect to the coordinatewise convergence on the space
$\R^{2\infty+2}$. Since $F$ is continuous, we get 
$F(\pi_N(t))\to F(\pi(t))$. That is, $F\circ\pi_N$ converges to
$F\circ\pi$ at any point $t\in\Hreg$. Since these functions are
uniformly bounded, it follows that 
$$
\int_{\Hreg}(F\circ\pi_N)(t)) M(dt)\,\to
\int_{\Hreg}(F\circ\pi)(t)) M(dt).
$$

Since $\pi_N(M)=P_N$ and $\pi(M)=P$,  
$$
\int_{\Hreg}(F\circ\pi_N)(t)) M(dt)=\langle F, P_N\rangle,\qquad
\int_{\Hreg}(F\circ\pi)(t)) M(dt)=\langle F, P\rangle.
$$
Consequently, $\langle F, P_N\rangle\,\to\,\langle F, P\rangle$. 
\qed
\enddemo

\head 6. The main result \endhead

Let $s\in\C$, $\Re s>-\tfrac12$. Consider
the Hua--Pickrell measure $\ms$. Let $P^{(s)}$ be its spectral measure and 
$\Cal P^{(s)}$ be the corresponding point process on $\R^*$, see \tht{0.8}.  

In this section we prove the following theorem which is our main
result. 

\proclaim{Theorem 6.1} The correlation functions of the process 
$\Cal P^{(s)}$ exist and coincide with the limit correlation functions from Theorem 2.1. 
\endproclaim

Let $X$ range over $\Hreg$. Recall that in \S5 we attached to $X$ two
monotone sequences $\{\al^+_i(X)\}$, $\{\al^-_i(X)\}$ and also, for any
$N=1,2,\dots$, two monotone sequences 
$$
\{a^+_{i,N}(X)=a^+_{i}(\la^{(N)}(X))\}, \qquad
\{a^-_{i,N}(X)=a^-_{i}(\la^{(N)}(X))\}.
$$
{}From these data we form point configurations 
$$
\Cal C(X)=\{\al^+_i(X)\}\sqcup\{-\al^-_i(X)\}, \qquad
\Cal C_N(X)=\{a^+_{i,N}(X)\}\sqcup\{-a^-_{i,N}(X)\},
$$
where we omit the zero coordinates.

Let $M$ be a $U(\infty)$-invariant probability measure on $H$. We
restrict $M$ to $\Hreg$, which is a subset of full measure, and view
$(\Hreg,M)$ as a probability space. Then any quantity depending on
$X$ becomes a random variable.

Let $P$ be the spectral measure of $M$ and let $P_N$ be the
finite--dimensional measures defined in Theorem 5.3. Recall that $P_N$
approximate $P$ as $N\to\infty$. 

Let $\Cal P_N$ and $\Cal P$ be the point processes on $\R^*$
corresponding to $P_N$ and $P$, respectively. We may view $\Cal P_N$
and $\Cal P$ as the random point configurations $\Cal C_N(X)$ and
$\Cal C(X)$, where $X$ is viewed as the point of the probability
space $(\Hreg,M)$.  

By $\rho^{(N)}_k$ and $\rho_k$ we denote the $k$th correlation
measures of the processes $\Cal P_N$ and $\Cal P$, respectively. Note
that the very existence of the measures $\rho_k$ is not evident. 

For a compact set $A\subset\R^*$ we set
$$
\Cal N_{A,N}(X)=\operatorname{Card}(\Cal C_N(X)\cap A), \qquad
\Cal N_A(X)=\operatorname{Card}(\Cal C(X)\cap A).
$$
These are random variables. 

We know that for any fixed $X$ and for any index $i=1,2,\dots$,
$a^\pm_{i,N}(X)$ tends to $\al^\pm_i(X)$ as $N\to\infty$. We would
like to conclude from this that $\rho^{(N)}_k$ converges to $\rho_k$
as $N\to\infty$. The next lemma says that, under a reasonable
technical assumption, this is indeed true. 

\proclaim{Lemma  6.2} Assume that for any compact set $A\subset \R^*$
there exist uniform in $N$ estimates 
$$
\E[\Cal N_{A,N}^l]\le C_l\,, \qquad l=1,2,\dots,
$$
where the symbol $\E$ stands for the expectation.

Then for any $k=1,2,\dots$, the correlation measure $\rho_k$ exists and
coincides with the weak limit of the measures $\rho^{(N)}_k$ as 
$N\to\infty$. The limit is understood in the following sense: for any
continuous compactly supported function $F$ on $(\R^*)^k$ 
$$
\lim_{N\to\infty}\langle F,\rho^{(N)}_k\rangle=
\langle F, \rho_k\rangle.
$$
\endproclaim

\demo{Proof} Fix a continuous compactly supported function $F$ on
$(\R^*)^k$. It will be convenient to assume that $F$ is nonnegative
(this does not mean any loss of generality). Introduce random variables $f$
and $f_N$ as follows: 
$$
f(X)=\sum_{x_1,\dots,x_k\in \Cal C(X)}F(x_1,\dots,x_k), \qquad
f_N(X)=\sum_{x_1,\dots,x_k\in \Cal C_N(X)}F(x_1,\dots,x_k), \tag6.1
$$
summed over ordered $k$-tuples of points with pairwise distinct
labels. Any such sum is actually finite because $F$ is compactly
supported and the point configurations are locally finite.

By the definition of the correlation measures,
$$
\langle F, \rho_k\rangle=\E[f], \qquad
\langle F, \rho^{(N)}_k\rangle=\E[f_N].
$$
The correlation measure $\rho_k$ exists if $\E[f]$ is finite for any
$f$ as above, see, e.g., \cite{Len}. 

Thus, we have to prove that $\E[f_N]\to\E[f]<\infty$ as $N\to\infty$. By a
general theorem (see \cite{Shir, ch. II, \S6, Theorem 4}), it
suffices to check the following two conditions:

{\it Condition 1.\/} $f_N(t)\to f(X)$ for any $X\in\Hreg$.

{\it Condition 2.\/} The random variables $f_N$ are uniformly
integrable, that is,
$$
\sup_N\int_{\{X\mid f_N(X)\ge c\}}f_N(X)M(dX)\to 0, \quad
\text{as $c\to+\infty$.}
$$

Let us check Condition 1. This condition does not depend on $M$, it
is a simple consequence of the regularity property. Indeed, let us
fix $X\in\Hreg$. For any 
$\ep>0$ set $\R^\ep=\R\setminus(-\ep,\ep)$.
Choose $\ep$ so small that the function $F$ is supported by
$(\R^\ep)^k$. Fix $j$ so large that $\al^\pm_j(X)<\ep$. Since
$a^\pm_{j,N}(X)\to \al^\pm_j(X)$, we 
have $a^\pm_{N,j}<\ep$ for all $N$ large enough. By monotonicity, the
same inequality holds for the indices $j+1,j+2,\dots$ as well. 

Recall that each point $x\in\Cal C_N(X)$ has the form
$x=a^+_{i,N}(X)$ or $x=-a^-_{i,N}(X)$ for a certain index $i$. It
follows that in the sums \tht{6.1}, only the points 
with indices $i=1,\dots,j-1$ may really contribute. Then, using the
continuity of $F$ we conclude that $f_N(X)\to f(X)$.  

Let us check Condition 2. Choose a compact set $A$ such that $F$ is
supported by $A^k$. The supremum of $F$ (let us denote it by 
$\sup F$) is finite. We have  
$$
f_N(t)\le \sup F\cdot \Cal N_{A,N}(X)(\Cal N_{A,N}(X)-1)
\dots(\Cal N_{A,N}(X)-k+1)
\le \sup F\cdot(\Cal N_{A,N}(X))^k.
$$
Therefore, the random variables $f_N$ are uniformly integrable
provided that this is true for the random variables 
$(\Cal N_{A,N})^k$ for any fixed $k$. But the latter fact follows from the
assumption of the theorem and Chebyshev's inequality. \qed
\enddemo

Assume that $\Cal P_N$ is a determinantal process given by a symmetric
nonnegative integral operator $K_N$ on $\R^*$. That is, the
correlation functions have determinantal form with the kernel $K_N$.
For a compact set 
$A\subset\R^*$ we denote by $K_{A,N}$ the restriction of the kernel
$K_N$ to $A$.  

\proclaim{Lemma 6.3} Assume that for any compact set $A\subset\R^*$
we have an estimate $\tr K_{A,N}\le \const$, where the constant does
not depend on $N$. Then the assumption of Lemma 6.2 is satisfied. 
\endproclaim

\demo{Proof} Instead of ordinary moments we can deal with factorial
moments. Given $l=1,2,\dots$, the $l$th factorial moment of
$\Cal N_{A,N}$ is equal to 
$$
\rho^{(N)}_l(A^l)=\int_{A^l}\det [K_{A,N}(x_i,x_j)]_{1\le i,j\le l}\,
dx_1\dots dx_l=l!\tr(\wedge^l K_{A,N}).
$$
Since $K_{A,N}$ is nonnegative, we have 
$$
\tr(\wedge^l K_{A,N})\le\tr(\otimes^l K_{A,N})=(\tr(K_{A,N}))^l.
$$
This concludes the proof, because we have a uniform bound for the
traces by the assumption. \qed
\enddemo

\demo{Proof of Theorem 6.1} Take $M=\ms$ and denote the correlation
measure $\rho^{(N)}_k$ by $\rho^{(s,N)}_k$. The latter measure is
calculated in \S1: it coincides with a scaling of the
$k$th correlation function $\boldsymbol\rho^{(s,N)}_k(x_1,\dots,x_k)$
for the $N$th pseudo--Jacobi ensemble. In terms of the corresponding
correlation functions, 
$$
\rho^{(s,N)}_k(x_1,\dots,x_k)
=N^k\boldsymbol\rho^{(s,N)}_k(Nx_1,\dots,Nx_k), \qquad
x_1,\dots,x_k\in\R^*.
$$

By Theorem 2.1, for each $k=1,2,\dots$, there exists a limit
$$
\lim_{N\to\infty}\rho^{(s,N)}_k(x_1,\dots,x_k)
=\rho^{(s,\infty)}_k(x_1,\dots,x_k), \tag6.2
$$
uniformly on compact subsets in $(\R^*)^k$. Moreover, the correlation
functions have determinantal form. It follows that the assumptions of
Lemma 6.3 are satisfied (indeed, $\tr K_{A,N}$ is simply the
integral of the first correlation function $\rho^{(s,N)}_1(x)$ over
$A$). Consequently, we may apply Lemma 6.2. By this lemma, the
correlation measures of the process $\Cal P^{(s)}$ exist and coincide
with limits of the measures $\rho^{(s,N)}_k$ as $N\to\infty$.
Therefore, they are nothing else than the measures
$\rho^{(s,\infty)}_k$ defined by the limit correlation functions 
\tht{6.2}. \qed
\enddemo

\head 7. Vanishing of the parameter $\ga_2$ \endhead

In this section we show that the parameter $\ga_2$ which is
responsible for the 
presence of the Gaussian component vanishes for the measure $m^{(0)}$. 

We start with a general result concerning an abstract 
$U(\infty)$-invariant probability measure $M$. As in \S6, let 
$\Cal P_N$ and $\Cal P$ denote the corresponding point processes on $\R^*$,
and let $\rho^{(N)}_1$ and $\rho_1$ be their first correlation
measures. We assume that $\rho^{(N)}_1$ approach $\rho_1$, as
$N\to\infty$, in the sense of Lemma 6.2: 
$$
\langle G,\rho^{(N)}_1\rangle\to\langle G,\rho_1\rangle 
\qquad \text{for any $G\in C_0(\R^*)$,} \tag7.1
$$
where $C_0(\R^*)$ denotes the space of continuous functions with
compact support on $\R^*$. In \S6 we verified
that the condition \tht{7.1} holds when $M$ is a Hua--Pickrell measure. 

\proclaim{Proposition 7.1} Let $M$ satisfy the condition \tht{7.1}.
Further, assume that 
$$
\lim_{\ep\to0}\int_{-\ep}^\ep x^2\rho_1^{(N)}(dx)=0\qquad
\text{uniformly in $N$.} \tag7.2
$$

Then the spectral measure $P$ of the measure $M$ is concentrated on
the subset $\ga_2=0$ in $\Om$. 
\endproclaim

\demo{Comment} The density of the measure $\rho_1$ may have a
singularity at 0. For instance, when $M=m^{(0)}$, the density
function is proportional to $1/x^2$. The condition \tht{7.2} means 
that the densities of the measures $\rho_1^{(N)}$, multiplied by $x^2$,
are uniformly integrable about $x=0$. 
\enddemo 

We need a simple lemma.

\proclaim{Lemma 7.2} Assume we are given sequences
$$
a^+_{1,N}\ge a^+_{2,N}\ge\dots\ge0, \qquad
a^-_{1,N}\ge a^-_{2,N}\ge\dots\ge0, \qquad N=1,2,\dots,
$$
such that
$$
\lim_{N\to\infty}a^\pm_{i,N}=\al^\pm_i, \qquad i=1,2,\dots
$$
and 
$$
\lim_{N\to\infty}\sum_{i=1}^\infty((a^+_{i,N})^2+(a^-_{i,N})^2)
=\de<+\infty, \qquad N=1,2,\dots
$$
Further, let $F(x)$ be an arbitrary continuous function on
$\R_+$ such that 
$$
F(x)=x^2 \qquad \text{for $|x|<\ep$}
$$
with a certain $\ep>0$. 
Set $\ga_2=\de-\sum\limits_{i=1}^\infty((\al^+_i)^2+(\al^-_i)^2)$ and
note that $\ga_2\ge0$. 

Then we have 
$$
\lim_{N\to\infty}\sum_{i=1}^\infty(F(a^+_{i,N})+F(-a^-_{i,N}))=
\sum_{i=1}^\infty(F(\al^+_i)+F(-\al^-_i))+\ga_2.
$$
\endproclaim

\demo{Proof} Fix $k$ so large that $\al^+_{k+1}<\ep$,
$\al^-_{k+1}<\ep$. Then $a^+_{k+1,N}<\ep$, $a^-_{k+1,N}<\ep$ for
sufficiently large $N$ and, moreover, $a^+_{i,N}<\ep$, $a^-_{i,N}<\ep$
for all $i\ge k+1$ by monotonicity. Likewise, $\al^+_{i}<\ep$,
$\al^-_{i}<\ep$ for $i\ge k+1$. Therefore,
$$
F(\pm a^\pm_{i,N})=(a^\pm_{i,N})^2 \quad \text{(for large $N$)}, \quad
F(\pm \al^\pm_{i})=(\al^\pm_{i})^2, \qquad i\ge k+1.
$$
It follows that
$$
\sum_{i=1}^\infty (F(a^+_{i,N})+F(-a^-_{i,N}))
=\sum_{i=1}^k (F(a^+_{i,N})+F(-a^-_{i,N}))+
\sum_{i=k+1}^\infty ((a^+_{i,N})^2+(a^-_{i,N})^2)
$$
and similarly
$$
\sum_{i=1}^\infty (F(\al^+_{i})+F(-\al^-_{i}))
=\sum_{i=1}^k (F(\al^+_{i})+F(-\al^-_{i}))+
\sum_{i=k+1}^\infty ((\al^+_{i})^2+(\al^-_{i})^2)
$$
As $N\to\infty$, we have
$$
\sum_{i=1}^k (F(a^+_{i,N})+F(-a^-_{i,N}))\to
\sum_{i=1}^k (F(\al^+_i)+F(-\al^-_i)), 
$$
by continuity of $F$, and
$$
\sum_{i=k+1}^\infty((a^+_{i,N})^2+(a^-_{i,N})^2)\to 
\sum_{i=k+1}^\infty((\al^+_i)^2+(\al^-_i)^2)+\ga_2,
$$
by the assumption of the lemma. This conludes the proof. \qed
\enddemo

\demo{Proof of Proposition 7.1} Let $X$ range over $\Hreg$. Recall
the notation $a^\pm_{i,N}(X)$ and
$\al^\pm_i(X)$ introduced in \S5 and in the beginning of $\S6$. Let
$\ga_2(X)$ denote the value of the parameter $\ga_2$ at the point
$\pi(X)\in\Om$, where $\pi\:\Hreg\to\Om$ is the projection defined in
\S5. Our aim is to prove that $\ga_2(X)=0$ almost everywhere with
respect to the measure $M$. This implies the claim of the proposition. 

Fix a continuous function
$F(x)\ge0$, with compact support on $\R$ and such that $F(x)=x^2$
near 0.  For any $X\in\Hreg$ set
$$
\gather
\varphi_N(X)=\sum_{i=1}^\infty
(F(a^+_{i,N}(X))+F(-a^-_{i,N}(X))), \\
\varphi_\infty(X)=\sum_{i=1}^\infty
(F(\al^+_{i}(X))+F(-\al^-_{i}(X))).
\endgather
$$
Applying Lemma 7.2 to the sequences
$a^\pm_{i,N}=a^\pm_{i,N}(X)$ and $\al^\pm_{i}=\al^\pm_{i}(X)$, we get
$$
\varphi_N(X)\to\varphi_\infty(X)+\ga_2(X), \qquad
X\in\Hreg\,.
$$

The functions $\varphi_N(X)$, $\varphi_\infty(X)$,
$\ga_2(X)$ are all nonnegative Borel functions. 
By Fatou's lemma (see, e.g., \cite{Shir, ch. II, \S6, Theorem 2}), 
$$
\liminf_{N\to\infty} \int_{t\in\T_\reg}\varphi_N(X)M(dX)\ge
\int_{X\in\Hreg}\varphi_\infty(X)M(dX)
+\int_{X\in\Hreg}\ga_2(X)M(dX). 
$$

Recall that in the beginning of \S6 we introduced the point
configurations $\Cal C_N(X)$ associated with an arbitrary
$X\in\Hreg$. We have 
$$
\varphi_N(X)=
\sum_{i=1}^\infty(F(a^+_{i,N}(X))+F(-a^-_{i,N}(X))
=\sum_{x\in \Cal C_N(X)}F(x),
$$
so that
$$
\int_{X\in\Hreg}\varphi_N(X)M(dX)
=\langle F,\rho^{(N)}_1\rangle.
$$
Likewise,
$$
\int_{X\in\Hreg}\varphi_\infty(X)M(dX)
=\langle F,\rho_1\rangle.
$$
Therefore,
$$
\liminf_{N\to\infty}\langle F,\rho^{(N)}_1\rangle\ge
\langle F,\rho_1\rangle+\int_{X\in\Hreg}\ga_2(X)M(dX). \tag7.3
$$

On the other hand, we will prove that
$$
\limsup_{N\to\infty}\langle F,\rho^{(N)}_1\rangle\le
\langle F,\rho_1\rangle. \tag7.4
$$
It will follow from \tht{7.3} and \tht{7.4} that $\ga_2(X)=0$ for
$M$-almost all $X$, because $\ga_2(X)\ge0$. 

To prove \tht{7.4} we represent $F(x)$, for an arbitrary $\ep>0$, in the
form 
$$
\gather
F(x)=F_\ep(x)+G_\ep(x),\\ 
\text{where}\,\, 0\le F_\ep(x)\le x^2,\quad
\operatorname{supp}F_\ep\subset[-\ep,\ep], \quad
F_\ep(x)=x^2 \,\, \text{near 0}, \quad
G_\ep\in C_0(\R^*).
\endgather
$$
Choosing $\ep$ small enough, we can make 
$\langle F_\ep,\rho^{(N)}_1\rangle$ arbitrarily small, uniformly 
in $N$, by virtue of the assumption \tht{7.2}. As for 
$\langle G_\ep,\rho^{(N)}_1\rangle$, it tends to $\langle
G_\ep,\rho_1\rangle$, by \tht{7.1}. This concludes the proof of
Proposition 7.1. \qed 
\enddemo

\proclaim{Theorem 7.3} The spectral measure of the measure $m^{(0)}$
is concentrated on the set $\ga_2=0$ in $\Om$.
\endproclaim

\demo{Proof} By virtue of Proposition 7.1, it suffices to verify the
condition \tht{7.2}. To do this, we use the fact that in our case the first
correlation function $\rho^{(N)}_1(x)=\rho^{(0,N)}_1(x)$ has a very
simple expression: 
$$
\rho^{(0,N)}_1(x)=\frac1\pi\,\frac{N^2}{1+N^2x^2}\,. \tag7.5
$$
The simplest way to check \tht{7.5} is to use the relationship to the
$N$th Dyson ensemble, where the first correlation function is identically 
equal to $N$. 

{}From \tht{7.5} and the trivial estimate $\frac{N^2x^2}{1+N^2x^2}\le1$
we readily conclude that the condition \tht{7.2} is indeed satisfied.
\qed
\enddemo

We expect that Theorem 7.3 holds for any Hua--Pickrell measure.

\head 8. Remarks and problems \endhead

\subhead Orthogonal polynomials on the circle \endsubhead
In this paper we deal with the pseudo--Jacobi ensemble \tht{1.1}
defined by the weight function \tht{1.4} on the real line. Instead of
this, one could work with the orthogonal polynomial ensemble
\tht{0.11}. Then we need orthogonal polynomials on the unit circle
$\Bbb T$ with the weight function 
$$
\gather
(1+u)^{\bar s}(1+\bar u)^s=2^a\,(1+\cos\varphi)^a\,e^{b\varphi}\,,\\
\text{where}  \quad u=e^{i\varphi}\in\Bbb T, \quad
-\pi<\varphi<\pi, \quad s=a+ib.
\endgather 
$$
For real $s$, the weight function depends only on 
$\Re u=\cos\varphi\in[-1,1]$. Then one can use a general trick described in 
\cite{Sz, \S11.5}. It allows one to express the polynomials on 
$\Bbb T$ in terms of two families of orthogonal polynomials on the
interval $[-1,1]$, which, in our case, turn out to be certain Jacobi
polynomials. This makes it possible to evaluate the
Christoffel--Darboux kernel and then pass to a limit as $N\to\infty$,
which leads to another derivation of Theorem 2.1 (for real $s$).
Perhaps, such an approach can be used for nonreal values of $s$ as
well.   

\subhead Painlev\'e V \endsubhead
Consider a kernel of the form 
$$
K(x',x'')=\frac{P(x')Q(x'')-Q(x')P(x'')}{x'-x''}\,,
$$
where the functions
$P$ and $Q$ satisfy a differential equation of the form
$$
\frac d{dx}\bmatrix
{P(x)}\\{Q(x)}\endbmatrix=A(x)\bmatrix{P(x)}\\{Q(x)}\endbmatrix
 $$
with a traceless rational 2$\times$2 matrix $A(x)$. Let $J$ be a
union of intervals inside the real line. Then the Fredholm
determinant $\det(1+K|_J)$ 
satisfies a certain system of partial differential equations with the
endpoints of $J$ 
regarded as variables, see \cite{TW}. In particular, when only one endpoint is
moving the corresponding ordinary differential equation often happens to be
one of the Painlev\'e equations. 

The kernel $K^{(s,\infty)}$ introduced in Theorem 2.1 is not an exception. In
particular, the function
$$
\sigma(t)=t\, \frac
{d\ln\det\left(1-K^{(s,\infty)}|_{(t^{-1},+\infty)}\right)}{dt}, \quad t>0, $$
satisfies a $\sigma$-version of the Painlev\'e V equation:
$$
-(t\sigma'')^2=(2(t\sigma'-\sigma)+(\sigma')^2+i(\bar{s}-s)\sigma')^2-
(\sigma')^2(\sigma'-2is)(\sigma'+2i\bar{s}),
$$
see \cite{BD} for details. Note that the approach of \cite{BD} is very
different from the machinery developed in \cite{TW}. 

\subhead Infinite measures \endsubhead
The construction of the Hua--Pickrell measures 
$\ms$, $\Re s>-\frac12$, given in \S3 can be
extended to arbitrary complex values of $s$. However, when 
$\Re s\le-\frac12$, $\ms$ ceases to be a probability measure and becomes   
an infinite measure. Its pushforward $\msN$ under the projection 
$\th_N: H\to H(N)$ makes sense only for sufficiently large values of
$N$. Specifically, $N$ must be strictly greater that $-2\Re s$. Then
the measure $\msN$ is defined, within a constant factor not
depending on $N$, by formula \tht{3.1}, where the factor $\const_N$
is subject to the recurrence relation
$$
\const_N=\const_{N-1}\,\frac{\pi^N\Ga(s+\bar s+N)}
{2^{s+\bar s+2N-2}\,\Ga(s+N)\Ga(\bar s+N)}\,.
$$

In other words, even if the measures $\msN$ are infinite, their
projective limit $\ms=\varprojlim\msN$ still exists. The reason
is that the fibers of the projection $H(N)\to H(N-1)$ have finite
mass with respect to the conditional measures provided that $N$ is large
enough. 

\proclaim{Problem} Define and study the spectral decomposition of
the infinite measures $\ms$, $\Re s\le-\frac12$.
\endproclaim

\subhead Representation--theoretic meaning of $U(\infty)$-invariant
measures on $H$ \endsubhead
Let $G(N)=U(N)\ltimes H(N)$ be the semidirect product of the group
$U(N)$ acting on the additive group $H(N)$ by conjugations.
Similarly, set 
$$
G=U(\infty)\ltimes H(\infty)=\varinjlim G(N).
$$ 
The groups $G(N)$ are examples of the so--called Cartan motion
groups, and the group $G$ is an infinite--dimensional version of the
groups $G(N)$. 

A unitary representation $T$ of the group $G$ is called {\it
spherical\/} if it possesses a cyclic unit vector $\xi$ which is invariant
with respect to the subgroup $U(\infty)\subset G$. There is a one--to--one 
correspondence between the classes of equivalence
of the pairs $(T,\xi)$ and the $U(\infty)$--invariant probability
Borel measures $M$ on $H$. Given $M$, the representation $T$ can be realized in
the Hilbert space $L^2(H,M)$. Elements $U\in U(\infty)$
and $A\in H(\infty)$ act on functions $f\in L^2(H,M)$ as follows
$$
(T(U)f)(X)=f(U^{-1}XU), \quad
(T(A)f)(X)=e^{i\tr(AX)}f(X), \qquad X\in H.
$$
In this realization, $\xi$ is the constant function 1. 

Consider the matrix 
coefficient $\varphi(g)=(T(g)\xi,\xi)$, called the {\it spherical
function.\/} Since $\varphi$ is $U(\infty)$--biinvariant, the
function $\varphi\mid_{H(\infty)}$, the restriction of $\varphi$ to
the subgroup $H(\infty)\subset G$, is a $U(\infty)$--invariant
positive definite normalized function on $H(\infty)$. It follows that
$\varphi\mid_{H(\infty)}$ coincides with the Fourier transform
\tht{4.1} of the $U(\infty)$--invariant probability Borel measure $M$.

Under the correspondence $(T,\xi)\leftrightarrow M$, ergodicity of $M$ is
equivalent to irreducibility of $T$. Note also that for an
irreducible spherical representation $T$, the vector $\xi$ is unique
(within a scalar multiple), so that the function $\varphi$ is an
invariant of $T$. 

Thus, irreducible spherical representations of the group
$G=U(\infty)\ltimes H(\infty)$ are parametrized by ergodic measures
on $H$. For more details about representations of the group $G$, see
\cite{Ol2}, \cite{Pi2}. 

\subhead The graph of spectra \endsubhead
Recall that by $\SS_N$ we denoted the subset of $\R^N$ formed by
vectors $\la$ with  weakly decreasing coordinates. For
$\mu\in\SS_{N-1}$ and $\la\in\SS_N$ we write $\mu\prec\la$ if the
coordinates of $\la$ and $\mu$ interlace:
$$
\la_1\ge\mu_1\ge\la_2\ge\dots\ge\la_{N-1}\ge\mu_{N-1}\ge\la_N\,.
$$
We set
$$
q_{N-1,N}(\mu,\la)=
\cases \prod\limits_{1\le i<j\le N-1}(\mu_i-\mu_j)/
\prod\limits_{1\le k<l\le N}(\la_k-\la_l), & \text{if $\mu\prec\la$,}\\
0, & \text{otherwise.}\endcases
$$
Note that for any $\la\in\SS_N$
$$
\int_{\SS_{N-1}}q_{N-1,N}(\mu,\la)d\mu=1, \qquad
d\mu=d\mu_1\dots d\mu_{N-1}\,.
$$

Let $M$ be an arbitrary $U(\infty)$--invariant probability Borel
measures and $P_N$ be the radial part of the
measure $\th_N(M)$ (this is a probability measure on $\SS_N$). Then
the measures $P_1,P_2,\dots$ satisfy the following consistency
relation: 
$$
\int_{\SS_N}q_{N-1,N}(\mu,\la)P_N(d\la)=
\text{the density of $P_{N-1}$ at $\mu$.}
$$
Conversely, if a sequence $\{P_N\}$ of probability measures satisfies
the above consistency relation for each pair of adjacent indices then
this sequence comes from a certain measure $M$.

Introduce the set $\Cal T$ formed by all infinite sequences 
$$
\tau=(\tau^{(1)}\prec\tau^{(2)}\prec\dots), \qquad \tau^{(N)}\in\SS_N\,.
$$
Consider the probability measures $\wt P$ on $\Cal T$ with the following
property: for each $N=2,3,\dots$, the probability that
$\tau^{(N-1)}$ lies in an infinitesimal region $d\mu$ about a point
$\mu\in\SS_{N-1}$ conditional that $\tau^{(N)}=\la$, is 
$q_{N-1,N}(\mu,\la)d\mu$. Any such measure $\wt P$ is uniquely
determined by a sequence $\{P_N\}$ satisfying the consistency
relations. Thus, the measures $\wt P$ bijectively
correspond to $U(\infty)$--invariant probability measures $M$ on $H$. 

We call the collection of sets $\{\SS_N\}$ together with the
functions $q_{N-1,N}(\mu,\la)$ the {\it graph of spectra.\/} This
term was suggested by Sergei Kerov. According to the philosophy of
\cite{VK} we call the functions $q_{N-1,N}(\mu,\la)$ the {\it
cotransition functions\/} of the graph of spectra. Here the term
``graph'' should not be understood literally, it only hints at a
similarity with some ``branching graphs'' like the Young graph
\cite{VK} or the Gelfand--Tsetlin graph \cite{BO}. For instance, the
set $\Cal T$ is an analogue of the set of paths in a branching graph.
It can be shown that the graph of spectra can be obtained from the
Gelfand--Tsetlin graph via a scaling limit procedure. 

\subhead Projective limit of the spaces $U(N)$ \endsubhead
There exist projections (not group homomorphisms!) $U(N)\to U(N-1)$
which correspond, via the Cayley transform, to the projections
$H(N)\to H(N-1)$. This allows one to form the projective limit space
$\frak U=\varprojlim U(N)$. The space $\frak U$ admits a natural
two--sided action of the group $U(\infty)$. The space $H$ is embedded
into $\frak U$, and the measures $\ms$ are transferred to $\frak U$
via this embedding. The resulting measures on $\frak U$ are
quasiinvariant with respect to the two--sided action of $U(\infty)$.
This makes it possible to construct for the group $U(\infty)$ analogs
of the biregular representation. See \cite{Ner2}, \cite{Ol5} for more detail.

\subhead Analogy with the infinite symmetric group and the
Poisson--Dirichlet distributions \endsubhead
The construction of the space $\frak U$ mentioned above is parallel
to the construction of the space $\varprojlim S(n)$ of virtual
permutations, see \cite{KOV}. Here $S(n)$ denotes the symmetric group
of degree $n$. The family of  theHua--Pickrell measures 
should be viewed as a counterpart of a family $\{\mu_t\}_{t>0}$ of
probability measures on the space of virtual permutations, see
\cite{KOV}. The Hua--Pickrell measures play the same role in harmonic
analysis on the group $U(\infty)$ as the measures $\mu_t$ do in
harmonic analysis on the infinite symmetric group $S(\infty)$. The
decomposition of the measures $\mu_t$ on ergodic components is
described by the Poisson--Dirichlet distributions. These are
remarkable probability measures on an infinite--dimensional simplex
(see \cite{Kin}), which were studied by many authors. Thus, the
spectral measures $P^{(s)}$ may be viewed as counterparts of the
Poisson--Dirichlet distributions. 

\subhead Other examples of group actions \endsubhead
The action of the group $U(\infty)$ on the space $H$ examined in the
present paper is connected with a particular series of flat
symmetric spaces $\{G(N)/U(N)=H(N)\}_{N=1,2,\dots}$ which in turn is
related to a series of compact symmetric spaces: we mean the spaces
$U(N)$ with the action of the groups $U(N)\times U(N)$. There exist
in all 10 infinite series of compact symmetric spaces and related
flat spaces. With each such series, one can associate an
infinite--dimensional group action on a space of infinite matrices
(see, e.g., \cite{Pi2}) and a family of `Hua--Pickrell measures' on
that space depending on a real or complex parameter (see
\cite{Ner2}). We expect that the results of the present paper can be
carried over to this more general context.

\head 9. Appendix: existence and uniqueness of decomposition
on ergodic components \endhead

Let $\MM$ be the set of $U(\infty)$-invariant probability Borel
measures on $H$. We equip $\MM$ with the Borel structure generated by
the functions of the form $M\mapsto \langle F,M\rangle$, where $M$
ranges over $\MM$ and $F$ is an arbitrary bounded Borel function on
$H$. 

Let the symbol $\ex(\dots)$ denote the set of extreme points of a
convex set. Recall that elements of $\ex\MM$ are called ergodic
measures. 

\proclaim{Theorem 9.1} {\rm(i)} $\ex\MM$ is a Borel subset in $\MM$.

{(ii)} For any $M\in\MM$ there exists a probability Borel measure $P$
on $\ex\MM$ representing $M$, i.e.,
$$
\langle F,M\rangle=\int_{\Cal M\in\ex\MM}\langle F,\Cal M\rangle 
P(d\Cal M) \tag9.1
$$
for any bounded Borel function $F$ on $H$.

{(iii)} The measure $P$ is unique.
\endproclaim

There exist different ways to prove such results, in particular:

(i) Representation--theoretic techniques.  

(ii) Dynkin's theorem about boundaries of general Markov processes,
see \cite{Dyn} and the references therein.

(iii) Choquet's theorem about existence and uniqueness of
barycentric decomposition in compact metrizable convex sets which are
`Choquet simplices', see \cite{Ph}. 

In the first way, we reduce the problem to that of decomposing a
spherical representation of the Cartan motion group $G$ (see
\S8 above). Here we must adapt the classical desintegration theory for
representations of locally compact groups and $C^*$--algebras (see
\cite{Dix}) to groups which are not locally compact but are inductive
limits of locally compact groups (see \cite{Ol1, \S3.6}). A crucial
fact is that $(G, U(\infty))$ is a Gelfand pair in the sense of
\cite{Ol4, \S6}. 

In the second way, one should use the graph of spectra (see \S8) to reduce
Theorem 9.1 to Dynkin's theorem. 

Below we shall follow the third way.

\proclaim{Proposition 9.2 (Choquet's theorems)} Let $\frak A$ be
a convex subset of a locally convex topological vector space $E$.
Assume that $\frak A$ is compact and metrizable. 

{\rm(i)} $\ex\frak A$ is a Borel subset of $\frak A$ (more precisely,
a $G_\de$ subset).

{(ii)} For any $a\in\frak A$ there exists a probability Borel measure $P$
on $\ex\frak A$ representing $a$, i.e.,
$$
f(a)=\int_{b\in\ex\frak A}f(b) P(db) \tag9.2
$$
for any continuous linear functional $f$ on $E$.

{(iii)} The measure $P$ is unique if and only if the cone spanned by
$\frak A$ is a lattice.
\endproclaim

\demo{Proof} Claim (i) is an elementary fact, see \cite{Ph, Prop.
1.3}. Claims (ii) and (iii) are Choquet's theorems, see \cite{Ph,
sections 3 and 9}.  \qed 
\enddemo

We need one more general result.

\proclaim{Proposition 9.3} For any group action on a Borel space, the
cone of finite Borel measures is a lattice.
\endproclaim

\demo{Proof} See \cite{Ph, section 10}. \qed
\enddemo

By Proposition 9.3, the set $\MM$ satisfies the lattice condition,
from the last claim of Choquet's theorem. However, there
is no apparent way to make $\MM$ a compact space, which is the major
difficulty to apply Choquet's theorem. We surmount it by embedding
$\MM$ into a larger convex set to which Choquet's theorem is
applicable. Here we use an idea borrowed from the proof of Theorem
22.10 in \cite{Ol3} (see also section 6 in \cite{OkOl}).  

\demo{Proof of Theorem 9.1} For $N=1,2,\dots$ let $\MM_N$ denote the
set of $U(N)$-invariant probability Borel measures on $H(N)$ and let
$\wt\MM_N$ be the larger set formed by $U(N)$-invariant finite Borel
measures of total mass less or equal to 1.

Further, let $C_0(H(N))$ be the Banach space of continuous functions on
$H(N)$ vanishing at infinity, and let $E_N$ denote its dual space
equipped with the weak star topology. Using the natural pairing between
functions from $C_0(H(N))$ and finite measures, we embed $\wt\MM_N$
into $E_N$. Note that $\wt\MM_N$ is a compact metrizable space with
respect to the topology of $E_N$. 

For $N=2,3\dots$ let $\th_{N-1,N}$ denote the projection $H(N)\to
H(N-1)$ which consists in removing the $N$th row and column from a
$N\times N$ matrix. This projection sends $\wt\MM_N$ to
$\wt\MM_{N-1}$ and also sends $\MM_N$ to $\MM_{N-1}$. Moreover, $\MM$
coincides with the projective limit space $\varprojlim\MM_N$.

Note that the map 
$\th_{N-1,N}:\wt\MM_N\to\wt\MM_{N-1}$ is not continuous. The reason
is that the projection $H(N)\to H(N-1)$ is not a proper map. (To
illustrate this phenomenon, consider the projection of the plane
$\R^2$ onto its first coordinate axis. Take the Dirac measure at a
point on the second coordinate axis and move the point to infinity.
Then the measure will weakly converge to the zero measure, while its
projection will remain fixed.)  

However, the map $\th_{N-1,N}:\wt\MM_N\to\wt\MM_{N-1}$ possesses a
weaker property: it is semicontinuous from below. (This property
does not rely on the specific character of the projection 
$H(N)\to H(N-1)$, it holds for any continuous map between locally
compact spaces.) This implies that for any $N=2,3,\dots$ the set
$$
A_{N-1,N}=\{(M_{N-1},M_N)\in\wt\MM_{N-1}\times\wt\MM_N
\mid M_{N-1}\ge\th_{N-1,N}(M_N)\} \tag9.3
$$
is closed.

It is convenient to allow the index $N$ in \tht{9.3} take the value
$\{1\}$. To this end we define $H(0)$ as a one--point set. Then $\th_{0,1}$
projects $H(1)$ into a single point, the vector space $E_0$ is
identified with $\R$, $\wt\MM_0$ is the interval $[0,1]\in E_0$, and
$\MM_0$ is identified with 1.

Next, we take as $\frak A$ the subset of $E_0\times E_1\times\dots$
formed by infinite sequences $a=(M_0,M_1,\dots)$ such that $M_0=1$,
$M_N\in\wt\MM_N$ for $N=1,2,\dots$, and for any $N=1,2,\dots$, the
pair $(M_{N-1,N},M_N)$ belongs to the set $A_{N-1,N}$ defined in
\tht{9.1}. We remark that $\frak A$ is a convex compact metrizable
set. 

For any $N=0,1,2,\dots$ we define an embedding 
$\iota:\MM_N\to\frak A$ as follows: 
$$
\gather
\MM_N\ni M\mapsto a=(M_0,M_1,\dots,M_N,0,0,\dots),\\
M_N=M, \qquad M_{i-1}=\th_{i-1,i}(M_i), \quad i=N,\dots,1.
\endgather
$$
We also consider the embedding $\iota:\MM\to\frak A$ which comes from
the identification of $\MM$ with $\varprojlim\MM_N$. 

Now, we make the following crucial observation:

(*) {\it Any element $a\in\frak A$
can be written as a convex combination of certain elements 
$a_N\in\iota(\MM_N)$ and an element $a_\infty\in\iota(\MM)$. Moreover, this
representation is unique.\/} 

By Proposition 9.3, for any $N$, the cone in $E_N$ spanned by $\MM_N$
is a lattice, and the same is also true for $\MM$. Together with (*),
this implies that the cone generated by $\frak A$ is a lattice. Thus,
the set $\frak A$ satisfies all the assumptions of Choquet's theorem.
Applying this theorem, we get that any point $a\in\frak A$ is
uniquely represented by a probability measure $P$ on $\ex\frak A$.

On the other hand, (*) implies the following fact:

(**) {\it $\ex\frak A$ is the disjoint
union of the sets $\iota(\ex\MM), \iota(\ex\MM_0), \iota(\ex\MM_1),\dots$.}

Since $\ex\frak A$ is a Borel set by Choquet's theorem, and since all
the sets $\iota(\ex\MM_N)$ are evidently Borel sets, we 
conclude from (**) that $\iota(\ex\MM)\subset\frak A$ is a Borel set.

Next, we note that the Borel structure on $\MM$ coming from
its embedding to $\frak A$ coincides with its initial Borel
structure. Indeed, both structures are defined by functions on $\MM$ of the
form $M\mapsto\langle F,M\rangle$, the only difference consists in
the choice of a class $\{F\}$ of functions on the space $H$. In the
latter case, $F$ may be an arbitrary bounded Borel function, while
in the former case $F$ belongs to the smaller class of cylindrical
functions of the form $G\circ\th_N$ with $G\in C_0(H(N))$,
$N=1,2,\dots$. However, both classes clearly generate the same Borel
structure. 

This proves claim (i) of Theorem 9.1.

Further, it follows from (**) and the definition of the set $\frak
A$ that if $a\in\MM$ then its representing measure $P$ is concentrated 
on $\iota(\ex\MM)\subset\ex\frak A$. Comparing \tht{9.1} and
\tht{9.2} we get that \tht{9.1} holds for any cylindrical function of
the form $F=G\circ\th_N$ with $G\in C_0(H(N))$. But then it also
holds for any bounded Borel function on $H$, as required. 
\qed
\enddemo

Recall that we have an explicit description of
the set $\ex\MM$: it is pa\-ra\-me\-trized by the space $\Om$ (Proposition
4.1). The next claim, together with Theorem 9.1, is used in
Proposition 4.4 above:

\proclaim{Proposition 9.4} The `abstract' Borel
structure on $\ex\MM$, which comes from the standard Borel structures
on $\MM$, coincides with the `concrete' Borel structure, which comes
from the natural Borel structure on $\Om$ via the bijection
$\ex\MM\leftrightarrow\Om$.  
\endproclaim

\demo{Proof}
Let us show that for any bounded Borel function $f$, the
expression $\langle f,M^\om\rangle$ is a Borel function in
$\om\in\Om$. Indeed, it suffices to check this claim for
functions $f$ of the form $f(X)=e^{i\,\tr(AX)}$, where $A$ is an
arbitrary fixed matrix from $H(\infty)$. Further, without
loss of generality we may assume that $A$ is a diagonal matrix, and
then the claim follows from Proposition 4.1.

Consider the correspondence $\ex\MM\leftrightarrow\Om$ provided by
Proposition 4.1. We have just proved that $\Om\to\ex\MM$ is a Borel
map. Since both $\Om$ and $\ex\MM$ are standard Borel spaces, we may apply a
general result (see \cite{Ma, Theorem 3.2}) to conclude that our correspondence
is an isomorphism of Borel spaces.  
\qed
\enddemo

\vskip 2 true cm

{\smc A.~Borodin}: Department of Mathematics, The University of
Pennsylvania, Philadelphia, PA 19104-6395, U.S.A.  

E-mail address:
{\tt borodine\@math.upenn.edu}

{\smc G.~Olshanski}: Dobrushin Mathematics Laboratory, Institute for
Problems of Information Transmission, Bolshoy Karetny 19, 101447
Moscow GSP-4, RUSSIA.  

E-mail address: {\tt olsh\@iitp.ru,
olsh\@online.ru}

\enddocument
\end